\documentclass[preprint,10pt,letterpaper]{aastex}

\slugcomment{Accepted for publication in \textit{The Astronomical Journal}. \copyright\ 2011 The American Astronomical Society. All rights reserved.}

\begin{document}

\title{Stellar Activity in the Broad-Band Ultraviolet}

\author{K. Findeisen, L. Hillenbrand}
\affil{California Institute of Technology, MC 249-17, Pasadena, CA 91125}
\email{krzys@astro.caltech.edu, lah@astro.caltech.edu}

\and

\author{D. Soderblom}
\affil{Space Telescope Science Institute, Baltimore, MD 21218}
\email{drs@stsci.edu}

\begin{abstract}
The completion of the GALEX All-Sky Survey in the ultraviolet allows activity measurements to be acquired for many more stars than is possible with the limited sensitivity of ROSAT or the limited sky coverage of Chandra, XMM, or spectroscopic surveys for line emission in the optical or ultraviolet. 
We have explored the use of GALEX photometry as an activity indicator, using as a calibration sample stars within 50~pc, representing the field, and in selected nearby associations, representing the youngest stages of stellar evolution.
We present preliminary relations between UV flux and the optical activity indicator $R^\prime_\textrm{HK}$ and between UV flux and age. We demonstrate that far-UV (FUV, 1350-1780\AA) excess flux is roughly proportional to $R^\prime_\textrm{HK}$.
We also detect a correlation between near-UV (NUV, 1780-2830\AA) flux and activity or age, but the effect is much more subtle, particularly for stars older than than $\sim 0.5-1$~Gyr.
Both the FUV and NUV relations show large scatter, $\sim 0.2$~mag when predicting UV flux, $\sim 0.18$~dex when predicting $R^\prime_\textrm{HK}$, and $\sim 0.4$~dex when predicting age.
This scatter appears to be evenly split between observational errors in current state-of-the-art data and long-term activity variability in the sample stars.
\end{abstract}


\section{Introduction}

The highest regions of the solar atmosphere, above the visible light photosphere, are heated primarily by magnetic processes rather than by convection or radiative diffusion. The solar magnetic field is believed to be driven from the interface between the radiative and convective zones, although the details of this interface dynamo are not yet well understood \citep{solarmodel_review}. Since all Sun-like stars have a radiative and a convective zone, magnetic fields should be a generic feature of all Sun-like stars. As a proxy for magnetic fields associated with other stars, we can observe optical, ultraviolet, or X-ray emission from magnetically heated gas, as well as flares, starspots, and other phenomena associated in the Sun with magnetic fields. These phenomena are collectively called stellar activity.

Stellar activity among Sun-like stars decays on time scales of $10^{8-9}$~yr \citep{soderblom91, redyouth, chro_levels}, much faster than the nuclear-burning timescale. This rapid evolution is understood to be a consequence of magnetic braking slowing stellar rotation \citep[e.g.][]{wind_braking}, in particular differential rotation, and of differential rotation driving the generation of magnetic fields \citep{solarmodel_review}. As a star ages it rotates more slowly, and thus its surface magnetic flux declines. A star with a lower magnetic flux cannot transport energy into its upper atmosphere as efficiently, and so the star's activity decays with time.
Understanding the evolution of stellar activity with age is important because activity is one of the few age indicators that can be applied to main sequence stars. In addition, observing trends between stellar activity, age, rotation, mass, and metallicity can improve our understanding of the physics underlying activity in both other stars and the Sun.

Because stellar activity is not a single phenomenon, it has many observational indicators. The most frequently used are the flux in the Ca~II H and K lines at 3934,68\AA, the soft X-ray flux, the H$\alpha$ 6563\AA\ equivalent width, and the amplitude of a star's photometric variability. 
Of these, the Ca~II lines are the most widely used because they have a number of advantages. 
They are resonance lines in the blue optical region, in which the non-thermal energy emitted by the outer stellar atmosphere can compete with the underlying thermal spectrum for detection. 
Other activity indicators accessible from the ground such as H$\alpha$ or the Ca~II IR triplet at 8498, 8542, and 8662\AA\ have much lower contrast, at least in solar-type stars, and are less efficiently observed. Stars from 0.5 to 1.0 in $B-V$ (F8~V to K2~V spectral types) behave in similar fashion with respect to their Ca~II emission. Stars as early as F5~V are also convective, but more weakly so, and stars later than K2~V show a different activity distribution in the Ca~II H and K lines, including stronger emission than would be expected extrapolating from G~dwarfs \citep{vpsurvey}.
The final advantage Ca~II has over other activity indicators is that its measurements can be calibrated in terms of the Mount Wilson S-index, $S_\textrm{MW}$, based on the original work of \citet{vpsurvey}. \citet{rhkcalc} define the conversion of $S_\textrm{MW}$ into another standard index, $R^\prime_\textrm{HK}$, representing the fraction of the star's bolometric luminosity emitted in the Ca~II lines through chromospheric activity. These two papers together allow any Ca~II flux measurement to be placed on a standard system, at least in principle.

A star's ultraviolet emission lines, particularly the C~II 1335\AA, C~IV 1549\AA, and Mg~II 2795,02\AA\ features, have also been used as activity indicators. Like Ca~II, they are high-contrast resonance lines, with formation temperatures ranging from $\sim 10^4$~K for Mg~II to $\sim 10^5$~K for C~IV.
\citet{1985ApJ...293..551S} explored the relationship of UV emission to age among field stars, adopting ages based on lithium depletion calculations. They found that, between the ages of 1~Myr and 3~Gyr, UV lines that form at $\sim 10^4$~K decay with an e-folding timescale of 2.7~Gyr, while $\sim 10^5$~K lines decay faster, in 1.4~Gyr.
\citet{sunasastar} used a small sample of solar analogs to conclude that, between the ages of 100~Myr and 6.7~Gyr, the combined line and continuum flux in the 100-360\AA\ band decays as $\tau^{-1.20}$ and that 920-1180\AA\ flux decays as $\tau^{-0.85}$.
So far, work in the UV band has been limited by the need to use space-based observatories, most of which have been designed for spectroscopy and can observe only small samples of stars.

With the launch of the GALEX ultraviolet survey mission \citep{galexintro} and the release of the GALEX All-Sky Imaging Survey (AIS), it is possible to measure ultraviolet fluxes for many relatively bright stars ($9 \lesssim V \lesssim 16$ for G~stars) entirely from archived observations. Such fluxes could be used to carry out an all-sky study of stellar activity of nearby stars, with more sensitivity than allowed by ROSAT X-ray observations, which detect only $\sim 20\%$ of nearby stars having $0.5 \le B-V \le 1.0$ \citep{rosatefficiency}. Alternatively, GALEX fluxes provide a way to select candidate active stars for more detailed study. Such a selection has already been done for M~dwarfs \citep{galexmdwarfs}, where activity produces an unambiguous UV excess.

In this paper, we explore the use of UV photometry as an activity indicator for Sun-like stars. 
In Section~\ref{hipsample}, we compare UV flux to the standard Ca~II indicator, $R^\prime_\textrm{HK}$, using a volume-limited sample of stars. In Section~\ref{mgsample}, we use young moving group and cluster members to look for a direct relationship between UV flux and age. In both sections we present empirical relations that predict $R^\prime_\textrm{HK}$ or age on the basis of UV flux, or vice versa.
In Section~\ref{fittest}, we verify that these relations do not suffer from a number of systematic errors.
Finally, in Section~\ref{discuss} we discuss the implications of our results.

\section{UV Broadband Flux and $R^\prime_\textrm{HK}$}\label{hipsample}

We explored the relationship between a star's flux in the GALEX FUV and NUV bands and its activity as measured by the $R^\prime_\textrm{HK}$ chromospheric index using a uniform sample of spectroscopic data. In Section~\ref{hiprhkdata} we present the $R^\prime_\textrm{HK}$ measurements, and in Section~\ref{hipgxdata} the GALEX magnitudes. In Section~\ref{hipresults} we present the method we used to fit relations between the UV magnitude and $R^\prime_\textrm{HK}$, examine the results of the fit, and construct UV-excess indicators, $R^\prime_\textrm{UV}$, defined analogously to $R^\prime_\textrm{HK}$.

\subsection{$R^\prime_\textrm{HK}$ Data}\label{hiprhkdata}

A large-scale survey of HK emission in nearby G dwarfs was undertaken by one of us (D.R.S.) from 1998 through 2002. The survey was done primarily using the Coud\'e Feed telescope at Kitt Peak National Observatory to observe sources from the north pole to -40\degr\ declination. Camera~5 was used with grating RC250 in first order and with the F3KB~CCD. A slit width of 250~$\mu$m gave good efficiency for the minimum resolving power of 4000 needed for the measurement. Stars below -40\degr\ declination were observed at Cerro Tololo Interamerican Observatory using the R-C spectrograph on the 1.5~m telescope; the details of these observations are the same as those reported in \citet{1996AJ....111..439H}. CTIO observations included some stars north of -40\degr\ to ensure cross-calibration. At both observatories various Mount Wilson standard stars spanning a broad range of color and activity levels were observed every run. Typical S/N levels were about 100 in the HK line cores but ranged over about 50 to 200 given changing observing conditions and approximate exposure times.  The data were reduced using standard IRAF routines to take account of dark frames, bias frames, flat fields, and wavelength calibration exposures.

The $R^\prime_\textrm{HK}$ measurements are on the Mount Wilson system, defined by the
spectrograph built by \citet{vpsurvey} with triangular band-passes centered on each of the H and K lines of width of 1.09\AA\ and two 20\AA\ continuum bands on either side of the HK region.  These filters are used to measure $S_\textrm{MW}$, which is the ratio of the flux in the cores of the H and K lines to that in the continuum. 
A correction for the star's spectral type, using the star's $B-V$ color, is then made to yield $R_\textrm{HK}$, the ratio of flux in the cores of the H and K lines relative to the star's bolometric flux. A correction for photospheric flux in the HK bandpasses, also based on $B-V$, is then made to produce $R^\prime_\textrm{HK}$ \citep{rhkcalc}.

Several stars in the sample were observed multiple times; the average scatter in $\log{R^\prime_\textrm{HK}}$ for these stars is 0.0813~dex. Our estimate of the scatter is consistent with previously published estimates, which are typically 0.05~dex in the $S_\textrm{HK}$ index \citep{sindex_wright, sindex_isaacson}. The $S_\textrm{HK}$ index shows less scatter than the $R^\prime_\textrm{HK}$ index because its flux includes a constant photosphere contribution, but we can correct for this distinction by multiplying the 0.05~dex figure by the median ratio of the Ca~II index with and without the photosphere, $R_\textrm{HK}/R^\prime_\textrm{HK} \sim 1.7$. It follows that previous studies have seen scatter in $R^\prime_\textrm{HK}$ of $\sim 0.085$~dex, similar to our own figure. \citet{sindex_error} suggests the systematic measurement errors in $R^\prime_\textrm{HK}$ are $\sim 0.04$~dex, leaving $\sim 0.07$~dex of the scatter from intrinsic variability.\label{hkvariable}

The goal was to observe a sample consisting of all G dwarfs within 50~pc.
A star was considered to be a G dwarf if it fell within 1 magnitude of a solar-composition zero-age main sequence (ZAMS) according to astrometry and photometry from the Hipparcos Catalog. Specifically, the star had to have a $B-V$ color from 0.50 to 1.00, inclusive, a Hipparcos parallax of 20~mas or larger with no regard paid to parallax error, and an absolute V magnitude (determined from the V magnitude in the Hipparcos catalog and the Hipparcos parallax) below a line constructed to be 1.0 magnitude above a ZAMS. A close approximation to this is a straight line drawn from $M_V = 3$ at $B-V = 0.5$ to $M_V = 5.5$ at $B-V = 1.0$.
As one would expect, the sample is distributed isotropically across the sky, except for one region containing excess stellar surface density from the Hyades cluster.

The sample observed is not a pure one in several regards. Ideally one would like to be able to define and then observe a complete, volume-limited sample of stars. In reality several effects are taking place.
First, the Hipparcos sample is imperfect. The Hipparcos Input Catalog attempted to be magnitude-limited to a V magnitude near 9 (essentially the HD catalog). However, for technical reasons the mission could observe a maximum number of stars in any one region and that means that some fairly bright stars are missing and that the magnitude limit has a ragged and inconsistent edge. Neither of these effects is likely to bias our sample because they were applied for strictly technical reasons with little regard for source type.

Second, the overall magnitude limit of the HIC does impose a bias, both in age and multiplicity. The Sun would have $V \sim 9$ if it were 50~pc away. Solar mass and composition stars younger than the Sun are not included beyond 50~pc due to their lower luminosity. Lower-mass stars suffer a larger age bias with a distance horizon for completeness that gets lower fast for cooler stars. 
However, we have tested for and found no detectable bias in the mean parallax with stellar color (see Section~\ref{biastests}), suggesting that this is not a major effect.
Binary systems can be brighter than single stars and so are more likely to be included in the HIC.

Third, the sample is only as pure as the data used to define it. This is generally not a problem because the stars in the sample are bright and well-studied, but sometimes the colors or magnitudes may be wrong. Also, the Hipparcos parallaxes have errors, and so there are some stars in the sample that are further away than we think and perhaps a few with underestimated parallaxes.

Fourth, we found obvious non-dwarfs during the course of the observing program. As noted, we paid no regard to parallax error in constructing our sample. One consequence is that the sample contains more distant stars that are non-dwarfs and that have large parallax errors, appearing to place them within our domain. Some of these appear to be cool white dwarfs or reddened A stars. We removed these contaminants from the sample, leaving 2680 stars.

\subsection{GALEX Data}\label{hipgxdata}

GALEX has a lower spatial resolution ($\sim 5\arcsec$) than Hipparcos, allowing cases where flux from a single GALEX source must be arbitrarily divided between two optical sources. 
In addition, double stars unresolved by either GALEX or Hipparcos could show discrepant colors that would confuse the analysis. 
To avoid such systematics, we removed known binary stars that were too close for GALEX to resolve, or whose separation we could not determine, by dropping stars meeting the following criteria in the Hipparcos catalog\footnote{The official Hipparcos designations of the fields we used are as follows: proximity flag is H2, number of components is H58, multiple systems flag is H59, and component separation is H64.}:\label{binarydrop}
\begin{itemize}
\item Either the proximity flag is set, there is more than one stellar component in the Hipparcos entry, or the multiple systems flag is equal to C, G, or O, and
\item The component separation is less than 10\arcsec\ or not specified
\end{itemize}
Of the 2680 stars in the original sample, we removed 19\% as potentially unresolved binaries.

We searched for counterparts from the GALEX All-Sky Imaging Survey (AIS) GR5 data release\footnote{\url{http://galex.stsci.edu/GR4/} [sic]} for each star remaining in the $R^\prime_\textrm{HK}$ sample. Many of the stars lay outside the survey's sky coverage, either because they were bright enough to damage the GALEX detectors or because they were located too close to the Galactic plane, where the number of stars poses a similar danger.
We required that GALEX sources be within 7\arcsec\ of the Hipparcos position, corrected for proper motion. The large search radius, compared to the GALEX astrometric accuracy of 0.5\arcsec, was needed because the GALEX PSF is considerably degraded for bright stars. 
We also required that sources be within 0.5\degr\ of the field center to minimize uncertainties in the PSF and the detector response. We did not filter the GALEX sources by artifact flags, as the flags (such as ``proximity to a bright source'') are all either invalid or irrelevant for bright stars. After removing stars that were not covered by the AIS, we were left with 1204 stars, of which all but one were detected by GALEX. HIP~48141 was not detected because it was observed only in the FUV band, where we are not sensitive to cooler stars.

\begin{figure*}
\includegraphics[width=0.32\textwidth]{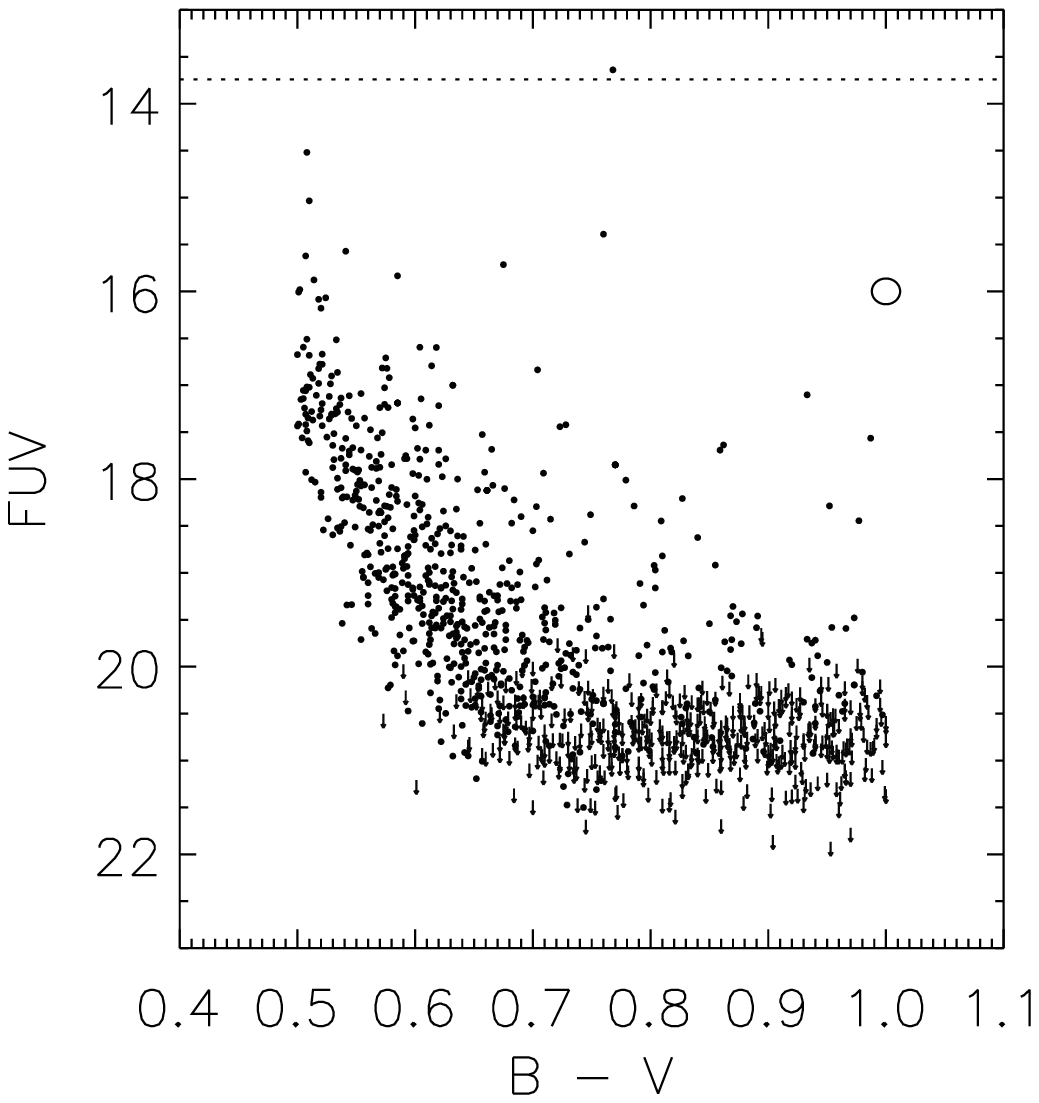}
\includegraphics[width=0.32\textwidth]{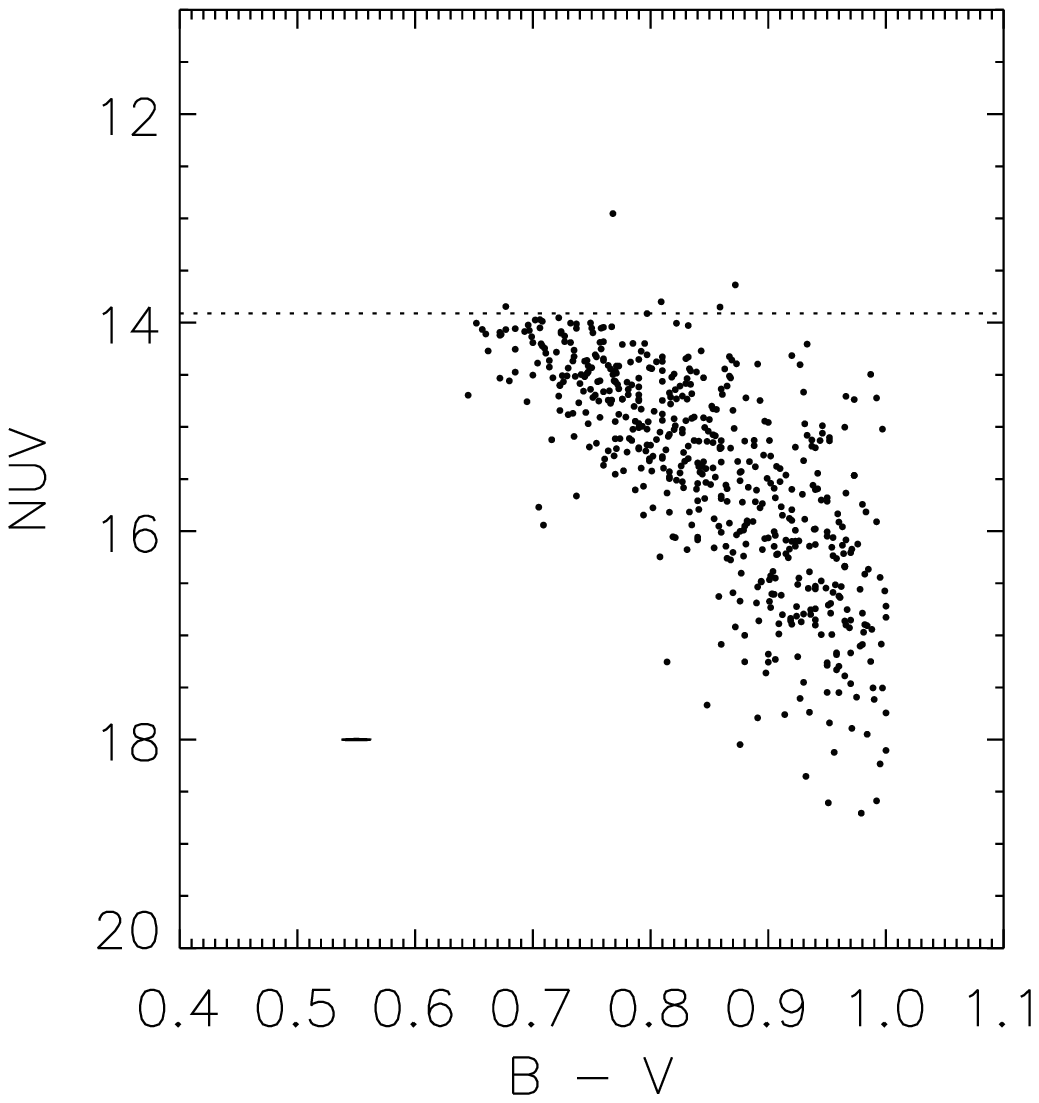}
\includegraphics[width=0.32\textwidth]{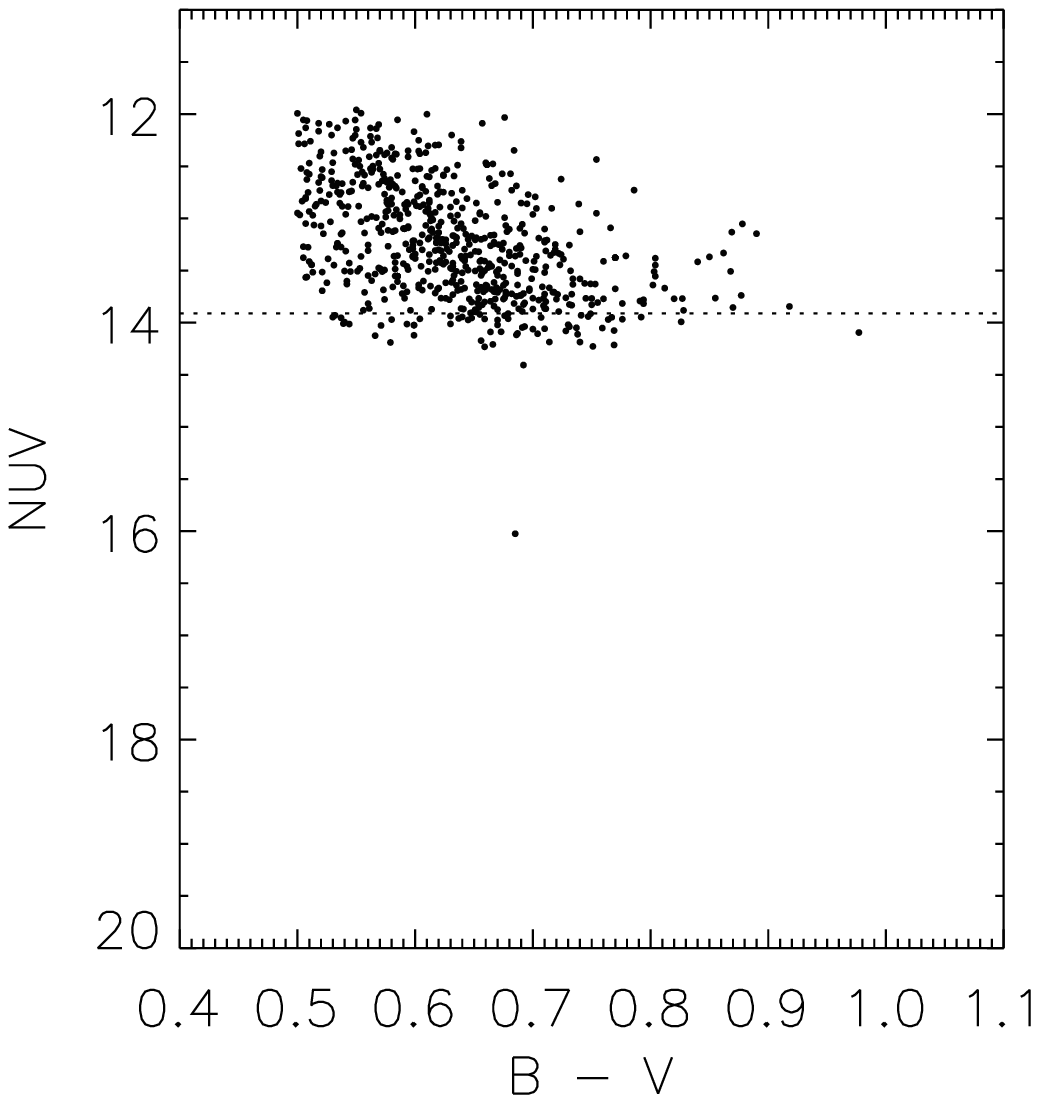}
\caption{$UV$ vs. $B-V$ diagrams of our volume-limited Hipparcos sample, showing the effect of saturation. The dotted line shows the magnitude at which nonlinearity effects introduce a 10\% or larger error \citep{galexcalib}. The ellipses in the upper right of the left panel and the lower left of the middle panel show the median errors in flux and color. The two right panels show the stars that did (center) and did not (right) pass the requirement that $V + 6.46 (B-V) > 12.8$. This cut effectively removed stars that saturated in the NUV, without introducing a bias toward stars that were underluminous in the NUV. Our FUV data does not saturate, so no cut was necessary.}\label{saturation_hip}
\end{figure*}

We used aperture photometry provided as part of the GALEX GR5 source catalog\footnote{\url{http://galex.stsci.edu/casjobs/}}, using a 17.25\arcsec\ radius. We added an 0.07~mag aperture correction in both the FUV and NUV, following \citet{galexcalib}. The median photometric error in the matched sample is 0.006~mag in the NUV, but only 0.13~mag in the FUV because many of the sources are faint in the FUV. We show the matched sample in Figure~\ref{saturation_hip}.

Many of the stars observed by GALEX were significantly saturated in the NUV. Since we cannot yet reliably correct the fluxes of heavily saturated GALEX sources, we could not use these stars in this paper. Instead of making a cut directly on NUV, which would bias the sample towards intrinsically UV-faint stars, we required that $V + 6.46 (B-V) > 12.8$. This cut, whose slope we derived from a linear fit to the $NUV-V$ and $B-V$ colors, ensured that most stars in our sample had $NUV > 13.9$, the magnitude below which saturation lowers the apparent flux by 10\% or more \citep{galexcalib}, but at the cost of removing most stars with $B-V \lesssim 0.7$ from our NUV sample. We illustrate the efficiency of the cut in Figure~\ref{saturation_hip}.

Many of the stars were not detected in the FUV. We constructed formal detection limits for these stars by finding the smallest flux that would give a $3\sigma$ flux measurement. We used an SNR cutoff, rather than trying to estimate the threshold below which the GALEX pipeline no longer identifies sources, to avoid any stochastic behavior in the source extraction algorithm at faint levels and to avoid including $S/N \sim 2$ sources whose formal flux measurement provides almost no information about the source flux. We modeled the photometric noise as $\sqrt{AB + F}$, where $A$ is the aperture area, $B$ is the sky background (obtained, via the source catalog, from the GALEX pipeline's smoothed background maps), and $F$ is the source flux. According to this model, to be detected at signal-to-noise ratio $S$ a star must have a flux
\begin{equation}\label{uplim}
F = \frac{1}{2} \left(S^2 + \sqrt{S^4 + 4 S^2 AB}\right)
\end{equation}
If a star was not detected, we quoted its flux as an upper limit at the value given by Equation~\ref{uplim} with $S = 3$. If a star was identified by the GALEX pipeline, but with a recorded flux less than that given by Equation~\ref{uplim}, we also reported it as an upper limit. This ensured that our detection limits were consistent across the entire sample, whether or not SExtractor was able to identify a source at lower SNR.

\subsection{Results}\label{hipresults}

\subsubsection{Color-Color Locus}

To illustrate the effect activity has on stellar SEDs, we plot in Figure~\ref{ccplots_hip} $UV-V$ vs. $B-V$ plots, with stars segregated by $R^\prime_\textrm{HK}$, and in Figure~\ref{uvacplots_hip} $UV-V$ vs. $R^\prime_\textrm{HK}$ plots, with stars segregated by $B-V$. Because all stars in the sample are within 50~pc, we assume no reddening.
A clear trend appears in the stars' $FUV-V$ color, in that more active stars with higher values of $R^\prime_\textrm{HK}$ consistently have bluer FUV-V colors. The $NUV-V$ color shows a similar, but much weaker trend; only for stars with $B-V \gtrsim 0.9$ can we clearly distinguish between active and inactive stars.
This behavior is not surprising. The photosphere contributes a larger fraction of the flux in the NUV than in the FUV over our investigated color range, $0.5 \le B-V \le 1.0$, as we expect from $T_{eff} \lesssim 6200$~K photospheres and as we confirm in Section~\ref{hipr}. Since any variation in stellar activity is diluted by the photosphere flux, we expect to see a more subtle effect in the NUV.

\subsubsection{Nonparametric Fits}

In this section we investigate whether UV and optical activity indicators trace one another.
Our goal is to convert between UV flux and $R^\prime_\textrm{HK}$, and use one as a predictor for the other. 
In the absence of a quantitative physical model for activity indicators in either the UV or the optical, we fit polynomial expansions to the data. To predict UV activity from Ca~II activity, we used an expectation-maximization (EM) algorithm as outlined by \citet{fitylimits} to incorporate upper limits to the FUV flux. Predicting $R^\prime_\textrm{HK}$ from UV flux was more challenging. Since we know of no algorithm for handling limits in independent variables, we were forced to restrict our FUV sample to avoid undetected stars. In analogy to our procedure for avoiding saturated NUV stars, we removed all stars from the FUV sample with $V + 12.03 (B-V) \ge 15.5$ when fitting for $R^\prime_\textrm{HK}$ as predicted by UV flux. As a result, while we predict FUV flux for all stars with $0.5 \le B-V \le 0.9$, our prediction of $R^\prime_\textrm{HK}$ \emph{from} FUV flux holds only if $B-V \lesssim 0.7$.

For both the predictor of UV flux and the predictor of $R^\prime_\textrm{HK}$, we chose the number of terms in the expansion that minimized the leave-one-out cross-validation score $CV = (1/n) \sum_D (y_i - \hat{f}_{(-i)}(x_i))^2$, where $D$ is the set of stars detected in the UV, $\hat{f}$ is our estimate of the true relation $f(X) = E(Y|X)$, and $\hat{f}_{(-i)}$ is the fit we would have obtained leaving out the data point $(x_i, y_i)$. Using $CV$ as a goodness-of-fit statistic, and minimizing it, approximately minimizes the expected mean square error $E\left[(1/n) \sum_D (\hat{f}(x_i) - f(x_i))^2\right]$, independent of the functional form of the fit or the error distribution around it \citep[e.g.]{crossvalidation}.

We recognize that the procedure of \citet{fitylimits} has a number of limitations. It cannot be extended to known measurement errors, and in fact implicitly assumes that the scatter is intrinsic to the system being studied. Unfortunately, to our knowledge there is no good algorithm that both avoids similar assumptions and applies to models more complex than a straight line. In light of such concerns, we test the robustness of our fits in Section~\ref{fittest}.

\def\Runit{(\log{R^\prime_\textrm{HK}}+4.5)}
\def\Bunit{(B-V-0.8)}
\def\Nunit{(NUV-V-7)}
\def\Funit{(FUV-V-12)}
We found the following fits between a star's $UV-V$ color and its $R^\prime_\textrm{HK}$ value. All four relations require a $B-V$ color to control for the dependence of $UV-V$ color on spectral type. After each equation we give the region of $UV-V$ vs. $B-V$ or $R^\prime_\textrm{HK}$ vs. $B-V$ parameter space where it is valid. 
\begin{eqnarray}
\label{fuvfit}
FUV-V & = & 12.30 - 3.95 \Runit - 2.94 \Runit^2 \nonumber \\
&&	{}+ 2.22 \Bunit - 9.11 \Bunit\Runit \nonumber \\
&&	{}- 12.0 \Bunit\Runit^2\nonumber \\
&&	{}- 17.6 \Bunit^2 - 3.9 \Bunit^2\Runit \nonumber \\
&&	{}- 24.3 \Bunit^2\Runit^2 \\ 
&&	(0.5 \le B-V \le 0.9 \textrm{ and } 1.7 (B-V) \le \log{R^\prime_\textrm{HK}} + 6.62 \textrm{ and } -5.2 \le \log{R^\prime_\textrm{HK}} \le -4.2) \nonumber \\ 
\nonumber \\
\label{facfit}
\log{R^\prime_\textrm{HK}} & = & -4.25 - 0.44 \Funit \nonumber \\
&&	{}+ 3.48 \Bunit - 0.35 \Bunit\Funit \\ 
&&	(0.5 \le B-V \le 0.7 \textrm{ and } 3.4 \le FUV-V - 12.0 (B-V) \le 5.2) \nonumber \\
\nonumber \\
\label{nuvfit}
NUV-V & = & 6.19 - 0.87 \Runit - 0.93 \Runit^2 \nonumber \\
&&	{}+ 6.54 \Bunit - 2.73 \Bunit\Runit \nonumber \\
&&	{}- 2.89 \Bunit\Runit^2 \\
&&	(0.7 \le B-V \le 1.0 \textrm{ and } -5.2 \le \log{R^\prime_\textrm{HK}} \le -4.2) \nonumber \\
\nonumber \\
\label{nacfit}
\log{R^\prime_\textrm{HK}} & = & -5.17 - 0.29 \Nunit + 0.64 \Nunit^2 \nonumber \\
&&	{}+ 0.28 \Nunit^3 + 4.06 \Bunit \nonumber \\
&&	{}- 2.68 \Bunit\Nunit \nonumber \\
&&	{}- 2.15 \Bunit\Nunit^2 \nonumber \\
&&	{}- 0.57 \Bunit\Nunit^3 \\
&&	(0.7 \le B-V \le 1.0 \textrm{ and } 0.7 \le NUV-V - 6.5 (B-V) \le 1.6) \nonumber
\end{eqnarray}
Like all physically unmotivated fits, Equations~\ref{fuvfit}-\ref{nacfit} are best thought of as interpolations over the region of parameter space where we have data, rather than as functions whose form has physical significance.

The RMS residuals around Equations~\ref{fuvfit}-\ref{nacfit} are 0.33~mag, 0.15~dex, 0.21~mag, and 0.18~dex, respectively. Propagating the formal photometric errors and the observed scatter in $\log{R^\prime_\textrm{HK}}$, the expected residuals around Equations~\ref{fuvfit}-\ref{nacfit} are 0.28~mag, 0.10~dex, 0.16~mag, and 0.28~dex, respectively. Since the error propagation makes a variety of assumptions, such as Gaussian errors, which are only approximately true, we do not expect a perfect match between the expected and actual residuals.

The expected residuals have a large contribution from the photosphere color $B-V$, because of the steep dependence on UV photosphere flux on effective temperature, and a comparable contribution from scatter in $R^\prime_\textrm{HK}$.
For Equations~\ref{fuvfit} and \ref{facfit}, scatter in $R^\prime_\textrm{HK}$ contributes $\sim$ 60\% of the error budget, with uncertainties in $FUV-V$ and $B-V$ contributing $\sim$ 30\% and $\sim$ 10\%, respectively. The expected residuals around Equation~\ref{nuvfit} are dominated by uncertainties in $B-V$ (84\%), plus a 14\% contribution from scatter in $R^\prime_\textrm{HK}$. The residuals around Equation~\ref{nacfit} are evenly split between uncertainties in $B-V$ and $R^\prime_\textrm{HK}$ (45\% and 54\%, respectively). 
If $\sim 75\%$ of the variance in the $R^\prime_\textrm{HK}$ measurements is from intrinsic variability (Section~\ref{hkvariable}), then variability in $R^\prime_\textrm{HK}$ accounts for 40-45\% of the residuals around Equations~\ref{fuvfit}, \ref{facfit}, and \ref{nacfit}.

\subsubsection{Normalized Excess Fluxes}\label{hipr}

Equations~\ref{fuvfit}-\ref{nacfit} are purely empirical results, unnormalized by the stellar photosphere and independent of any theoretical expectations of how much UV flux an active star will produce. However, it is often useful to divide the UV flux into a fixed photosphere component and an excess associated with stellar activity.
Therefore, we also used Kurucz photosphere models to calculate broadband UV excess activity indices $R^\prime_\textrm{UV}$, in analogy to the spectroscopic activity index $R^\prime_\textrm{HK}$, as follows.

The GALEX magnitudes are defined on the AB system \citep{abmags}, allowing us to estimate the total flux observed in the GALEX bands as
\begin{equation}
f_\textrm{UV} = (1 \textrm{ erg s$^{-1}$ cm$^{-2}$ Hz$^{-1}$}) \times 10^{-0.4(UV+48.60)} \times \Delta\nu_\textrm{UV}
\end{equation}
where $\Delta\nu$ is the frequency range corresponding to the wavelength range 1350-1780\AA\ (for FUV) or 1780-2830\AA\ (for NUV). 
The adopted bandpasses represent the wavelengths where the effective area falls to 10\% of its peak \citep{galexcalib}, rounded to 10\AA\ precision for notational convenience. While the bandpasses are somewhat arbitrary, we show in Section~\ref{colortests} that they make the fluxes we calculate insensitive to the color of the source at the $\sim 10\%$ level.

We found bolometric corrections and photosphere UV fluxes to each star in our sample, using the spectral type fitting method of \citet{mscolors}. We constrained the fits using all available photometry, namely the Hipparcos $B$ and $V$ magnitudes and $J$, $H$, and $K$ magnitudes from the Two-Micron All Sky Survey. Table~\ref{modelmags} summarizes the run of empirical magnitudes in these five bands and bolometric magnitudes along the main sequence, based on \citet{mscolors} Table~5. 
To fit photosphere UV fluxes we also needed UV magnitudes along the main sequence, which we obtained from solar-metallicity Kurucz ``ODFNEW'' models\footnote{\url{http://kurucz.harvard.edu/}} for each spectral type. 
We matched the Kurucz grid to each spectral type by using the masses, effective temperatures and bolometric magnitudes in \citet{mscolors} Table~5 to infer values of $\log{g}$ at each spectral type, then linearly interpolated Kurucz models to the $(T_\textrm{eff}, \log{g})$ pair. Since UV and optical magnitudes are normalized on two different systems, we used radii inferred from the bolometric magnitudes and effective temperatures in \citet{mscolors} Table~5 to place the model fluxes on an absolute scale.

We then calculated absolute $FUV$ and $NUV$ magnitudes for each interpolated model by integrating the photon flux over GALEX effective area curves provided by A. Gil~de~Paz, and dividing by photon fluxes integrated over a CALSPEC\footnote{\url{http://www.stsci.edu/hst/observatory/cdbs/calspec.html}} spectrum (version mod\_002) of LDS~749B, the primary GALEX calibrator \citep{galexcalib}. Working with photon fluxes allowed our synthetic magnitudes to automatically incorporate color corrections, which can grow up to $\sim 0.2$~mag for K stars. 
We also calculated absolute $B$ and $V$ magnitudes using response curves from \citet{bessellcurves}, a CALSPEC spectrum of Vega (stis\_005), and the Landolt magnitudes of $B = 0.02$, $V = 0.03$ for Vega \citep{landolt_vega}. This ensured that, while these $B$ and $V$ magnitudes could not be considered color-corrected in the same sense as the $FUV$ and $NUV$ magnitudes, all four magnitudes were calculated consistently.

When the $FUV$ and $NUV$ magnitudes calculated from each model were matched to the spectral type the model was intended to represent, they produced a curve that did not match our observed stellar locus in color-color plots, as shown in Figure~\ref{synthuvmags}. In particular, the predicted photosphere NUV magnitude was typically larger than the observed (photosphere plus activity) NUV magnitudes. Investigating, we found that our synthetic $V$ vs. $B-V$ relation closely matched the data, while the $B-V$ vs. $T_\textrm{eff}$ did not, suggesting that the problem lay in the effective temperature adopted for each spectral type. 
We 
created the UV magnitudes we list for each spectral type in Table~\ref{modelmags} by linearly interpolating the model $UV-V$ vs. $B-V$ relation to the $B-V$ observed for each spectral type, and adding the observed $V$ magnitude. Figure~\ref{synthuvmags} shows that these corrected magnitudes appear much more consistent with our observations, particularly in the NUV.
%

\label{rprimeuv}
Having constructed Table~\ref{modelmags}, we estimated bolometric, photosphere FUV, and photosphere NUV magnitudes for each star as described in \citet{mscolors}. We then calculated the activity indices
\begin{equation}
R^\prime_\textrm{FUV} = \frac{f_\textrm{FUV} - f^\textrm{phot}_\textrm{FUV}}{f_\textrm{bol}} 
\qquad 
R^\prime_\textrm{NUV} = \frac{f_\textrm{NUV} - f^\textrm{phot}_\textrm{NUV}}{f_\textrm{bol}} 
\end{equation}
analogous to $R^\prime_\textrm{HK}$. Here, $f_\textrm{UV}$ is the flux inferred from the observed FUV or NUV magnitude, as defined at the beginning of this section, $f^\textrm{phot}_\textrm{UV}$ is the flux inferred in the same way from the estimated photosphere FUV or NUV magnitude, and $f_\textrm{bol}$ is the estimated bolometric flux. 
We propagated errors on $R^\prime_\textrm{UV}$ neglecting any correlation between $f^\textrm{phot}_\textrm{UV}$ and $f_\textrm{bol}$; even if the two estimates were perfectly correlated, the correlation term would constitute only $\sim 2\%$ of the error budget. 
The median uncertainty in $\log{R^\prime_\textrm{FUV}}$ was 0.082~dex, $\sim 80\%$ of which was propagated from uncertainties in $f_\textrm{FUV}$. The uncertainty in $\log{R^\prime_\textrm{NUV}}$ was 0.100~dex, $\sim 95\%$ from uncertainties in $f^\textrm{phot}_\textrm{NUV}$.

We compare the $R^\prime_\textrm{UV}$ and $R^\prime_\textrm{HK}$ indices in Figure~\ref{activity2}.
$R^\prime_\textrm{NUV}$ shows no correlation with $R^\prime_\textrm{HK}$, and a mean of $-2.0 \times 10^{-5} \pm 3.1 \times 10^{-5}$, consistent with zero. Since we \emph{do} see a correlation between $NUV-V$ with $R^\prime_\textrm{HK}$, we believe our lack of a correlation for $R^\prime_\textrm{NUV}$ represents the uncertainties in the photospheric contribution $f^\textrm{fit}_\textrm{NUV}$ overwhelming any variation in NUV flux caused by activity.
$R^\prime_\textrm{FUV}$ is well correlated with $R^\prime_\textrm{HK}$. We fit a line to the logs of both activity indicators using the procedure of \citet{fitylimits}, getting the relation
\begin{equation}\label{hip_ruvfit}
\log{R^\prime_\textrm{FUV}} = (0.98 \pm 0.05) \log{R^\prime_\textrm{HK}} + (-0.53 \pm 0.25)
\end{equation}
We compare our estimates of $R^\prime_\textrm{FUV}$ to previous work on UV stellar activity in Section~\ref{rprimecompare}.

\section{UV Broadband Flux and Stellar Age}\label{mgsample}

The volume-limited Hipparcos sample of Section~\ref{hipsample} is uniform and assumed to be extinction-free, but because it consists of field stars we do not have precise age estimates for individual members, nor do we probe ages much below 625~Myr, the age of the Hyades. To extend the study to younger ages, and to explore the dependence of UV flux on age itself rather than on an empirical proxy for age, we constructed a second sample consisting of members of nearby clusters and moving groups. We present this sample and the matched GALEX data in Section~\ref{mgdata}. In Section~\ref{mgresults} we present our relations between UV magnitude and age. Since our procedures for the moving group sample parallel those we used for the $R^\prime_\textrm{HK}$ sample, we summarize them here and refer the reader to Section~\ref{hipsample} for details.

\subsection{Data}\label{mgdata}

We selected members of the TW~Hydrae (8~Myr old), $\beta$~Pictoris (12~Myr old), Tucana/Horologium (30~Myr old), and AB~Doradus (50~Myr old) moving groups from \citet{mgreview_old}, as well as members of the clusters Blanco~1 (120~Myr old) from \citet{blancomembers} and the Hyades (625~Myr old) from \citet{hyamembers}. All ages are from the same papers as the membership lists. 
We could use only two clusters as age benchmarks because the vast majority of nearby clusters, such as the Pleiades, IC~2602, Alpha Persei, and Coma Berenices, have members bright enough to trip the GALEX safety limits and so could not be observed. More distant clusters, such as M~67 and NGC~188, are too distant for GALEX to detect members cooler than late G~type, so these clusters would produce samples that have little overlap in spectral type with the nearer groups.
We did not restrict membership to stars with $0.5 \le B-V \le 1.0$, as in Section~\ref{hipsample}, so compared to the Hipparcos sample the moving group and cluster sample has many more low-mass stars and a handful of higher-mass ones.

As in the Hipparcos sample, we
removed known binary stars that were too close for GALEX to resolve, or whose separation we could not determine, 
to avoid systematics associated with blended sources.
Since the data set was inhomogeneous, the criteria for removing binary stars varied by group:
\begin{description}
\item [Hyades:] \citet{hyamembers} considered only Hipparcos stars, so we dropped stars following the same rule as in Section~\ref{binarydrop}:
	\begin{itemize}
	\item Either the proximity flag is set, there is more than one stellar component in the Hipparcos entry, or the multiple systems flag is equal to C, G, or O, and
	\item The component separation is less than 10\arcsec\ or not specified
	\end{itemize}
\item [Blanco 1:] we dropped stars flagged by \citet{blancomembers} as resolved double stars or spectroscopic binaries
\item [Moving groups:] we used the SIMBAD database to look up literature on each star, and dropped stars with companions within 10\arcsec.
\end{description}
Of the 425 stars in the original sample, we removed 26\% as possible unresolved binaries in GALEX observations.

Many moving group members, particularly the lower-mass stars, do not have high quality optical photometry. Instead, we used infrared photometry from the Two-Micron All Sky Survey (2MASS) for our analysis, even though using infrared rather than optical colors increases the scatter around our relations (see Appendix~\ref{bestcolors}).
Each moving group or cluster member had exactly one 2MASS counterpart within 3\arcsec, so the matching was straightforward. The median errors on the J, H, and K magnitudes in the matched sample were 0.02, 0.03, and 0.02~mag, respectively.

We searched for counterparts within 7\arcsec\ of each star in the GALEX All-Sky Imaging Survey (AIS) GR5 data release. As with the Hipparcos sample, many of the stars could not be observed by GALEX because of the observatory's brightness limits. GALEX coverage proved a much more serious restriction than for the Hipparcos sample. Only 99 of the 313 single stars in the sample were observed by GALEX. The most badly affected group was Blanco~1, where only 6 outlying stars were observed out of 49 single members; the more centrally located members were too close to bright stars. Of the 99 targets three (TWA~26, TWA~28, and HD~89744~B), all brown dwarfs, were undetected by GALEX.

We processed GALEX sources in the same way as in Section~\ref{hipgxdata}: we considered only sources within 0.5\degr\ of the field center and used corrected aperture photometry from SExtractor. We avoided NUV-saturated stars by only considering stars with $J + 7.85 (J-K) > 10.3$ for NUV fits, and we constructed upper limits for missing fluxes from the exposure time and the local sky background using Equation~\ref{uplim}. The median FUV error for detected sources was 0.14~mag, while the median NUV error was 0.01~mag.

\subsection{Results}\label{mgresults}

We plot in Figure~\ref{ccplots_mg} $UV-J$ vs. $J-K$ plots, with stars segregated by group or cluster membership, and in Figures~\ref{uvacplots_mg}-\ref{acuvplots_mg} $UV-J$ vs. age plots, with stars segregated by $J-K$. There is a clear correlation between $NUV-J$ and age, in the sense that younger stars have bluer $NUV-J$ colors at fixed $J-K$. The FUV data, on the other hand, are confused by the large number of non-detections. 
We believe we see a stronger trend between NUV flux and age here than we saw between NUV flux and $R^\prime_\textrm{HK}$ in Section~\ref{hipresults} because we are considering younger stars, whose activity levels drop more rapidly than in the volume-limited sample.

Although these stars are generally more distant than those presented in Section~\ref{hipsample}, extinction is still negligible. The most heavily reddened group in our sample, Blanco~1, has $E(B-V) = 0.016$ \citep{blancoreddening}, which implies $E(FUV-J) \sim E(NUV-J) \sim 0.1$ based on the \citet{av_curve} reddening law. Since both the trends and the scatter in Figures~\ref{ccplots_mg} and \ref{uvacplots_mg} are much larger than 0.1~mag, we ignore foreground extinction.

We fit polynomial curves to the NUV data following the same procedure as in Section~\ref{hipresults}. We also tried to fit the FUV data, but the results were quite poor -- because many of the stars in this sample were cooler or more distant than those in the Hipparcos sample, we did not have enough FUV detections to meaningfully constrain the relationship.

\def\Aunit{(\log{(\textrm{age/yr})}-8.0)}
\def\Junit{(J-K-0.6)}
\def\Nunit{(NUV-J-8)}
\def\Funit{(FUV-J-12)}
We found the following fits between a star's $NUV-J$ color and its age. All four relations require a $J-K$ color to control for the dependence of $NUV-J$ color on spectral type. After each equation we give the region of $NUV-J$ vs. $J-K$ or $\log{\textrm{age}}$ vs. $J-K$ parameter space where it is valid. 
\begin{eqnarray}
\label{nuvjfit}
NUV-J & = & 8.77 + 0.79 \Aunit + 8.91 \Junit \nonumber \\
&&	{}+ 0.87 \Junit\Aunit - 3.57 \Junit^2 \nonumber \\
&&	{}- 3.92\Junit^2\Aunit \\
&&	(0.4 \le J-K \le 0.9 \textrm{ and } 6.9 \le \log{(\textrm{age/yr})} \le 8.8) \nonumber \\
\nonumber \\
\label{nacjfit}
\log{(\textrm{age/yr})} & = & 7.53 + 0.80 \Nunit \nonumber \\
&&	{}- 7.69 \Junit + 0.12 \Junit\Nunit \\
&&	(0.4 \le J-K \le 0.9 \textrm{ and } 2.7 \le NUV-J - 7.84 (J-K) \le 5.3) \nonumber
\end{eqnarray}

Equations~\ref{nuvjfit} and \ref{nacjfit} have residuals of 0.46~mag and 0.39~dex, respectively.
The expected residuals around these two fits are 0.63~mag and 0.43~dex from propagated errors in $NUV-J$ and $J-K$. 
We did not consider scatter from UV variability, although the NUV magnitudes of M~stars at the ages of the Pleiades and Hyades vary by $\sim 1$~mag \citep{galexpleiades}, and one might expect variability to be detectable against the brighter photospheres of G and K stars as well.
Errors in $J-K$ dominate the expected residuals, accounting for 97\% of the variance. We do not consider the effect of age errors, since they are highly correlated (i.e. our age estimates for all members of a group are in error by the same amount) and to first order should affect only the fits and not the scatter around the fits. As with the Hipparcos sample, we can account for the size of the residuals from propagated errors alone, particularly from poor J and K photometry for stars saturated in 2MASS. 

We also used the procedure of \ref{hipr} to estimate bolometric, photosphere FUV, and photosphere NUV magnitudes for our moving group members using their 2MASS photometry but no $B$ or $V$ data. The estimates were possible only for the bluer stars (inferred spectral type K5 or hotter), as the limited temperature coverage of the Kurucz grids did not allow us to find photosphere UV fluxes for cooler stars. We found a median error in $\log{R^\prime_\textrm{FUV}}$ of 0.089~dex, split evenly between propagated uncertainties from $f_\textrm{FUV}$ and from $f^\textrm{phot}_\textrm{FUV}$. The median error in $\log{R^\prime_\textrm{NUV}}$ was 0.26~dex, 95\% of which came from uncertainties in $f^\textrm{phot}_\textrm{NUV}$. 

We plot $R^\prime_\textrm{UV}$ vs. age in Figure~\ref{activityage}. As in Figure~\ref{activity2}, $R^\prime_\textrm{FUV}$ shows a trend with age, albeit with several orders of magnitude of scatter. There is a hint of a trend in $R^\prime_\textrm{NUV}$ as well, but as in Section~\ref{hipr} the calculation of $R^\prime_\textrm{NUV}$ is dominated by noise.
We were able to fit the evolution of the FUV flux as
\begin{equation}\label{mg_ruvfit}
\log{R^\prime_\textrm{FUV}} = (-0.42 \pm 0.18) \log{\textrm{(age/yr)}} + (-1.27 \pm 1.50)
\end{equation}
Figure~10 suggests that the nonzero slope of this fit is a result of the two Hyades members with the lowest FUV flux, i.e. the two points in the lower right corner of Figure~\ref{activityage}. Fitting the evolution while ignoring these two points gives 
\begin{displaymath}
\log{R^\prime_\textrm{FUV}} = (-0.33 \pm 0.16) \log{\textrm{(age/yr)}} + (-1.95 \pm 1.32)
\end{displaymath}
Thus, our fit is not significantly affected by the presence or absence of these two points.

\section{Testing for Systematic Errors}\label{fittest}

The relations described by Equations~\ref{fuvfit}-\ref{nacfit} and \ref{nuvjfit}-\ref{nacjfit} and the $R^\prime_\textrm{UV}$ indices require a host of assumptions. In this section, we verify that our results do not depend on these assumptions. In Section~\ref{montecarlo} we perform a simulation to test whether our procedure for fitting upper limits to the UV fluxes bias the resulting fits. In Section~\ref{nonparatest} we explore whether changing the degrees of the polynomials in Equations~\ref{fuvfit}-\ref{nacfit} and \ref{nuvjfit}-\ref{nacjfit} significantly changes the colors, $R^\prime_\textrm{HK}$ indices, or ages predicted by the fits. Finally, in Section~\ref{colortests} we explore whether our $R^\prime_\textrm{UV}$ indices depend greatly on the color of the source.

\subsection{Testing the Effects of FUV Incompleteness}\label{montecarlo}

In Sections~\ref{hipresults} and \ref{mgresults}, we used an expectation-maximization (EM) algorithm presented by \citet{fitylimits} to fit the UV flux upper limits in our data. The algorithm assumes that the scatter in the data is Gaussian, uniform across the sample, and intrinsic to the system under study rather than the result of measurement error. All three assumptions are violated in our data. Here we show, based on Monte Carlo simulations, that these assumptions do not invalidate our fits.

\begin{figure*}
\includegraphics[width=0.48\textwidth]{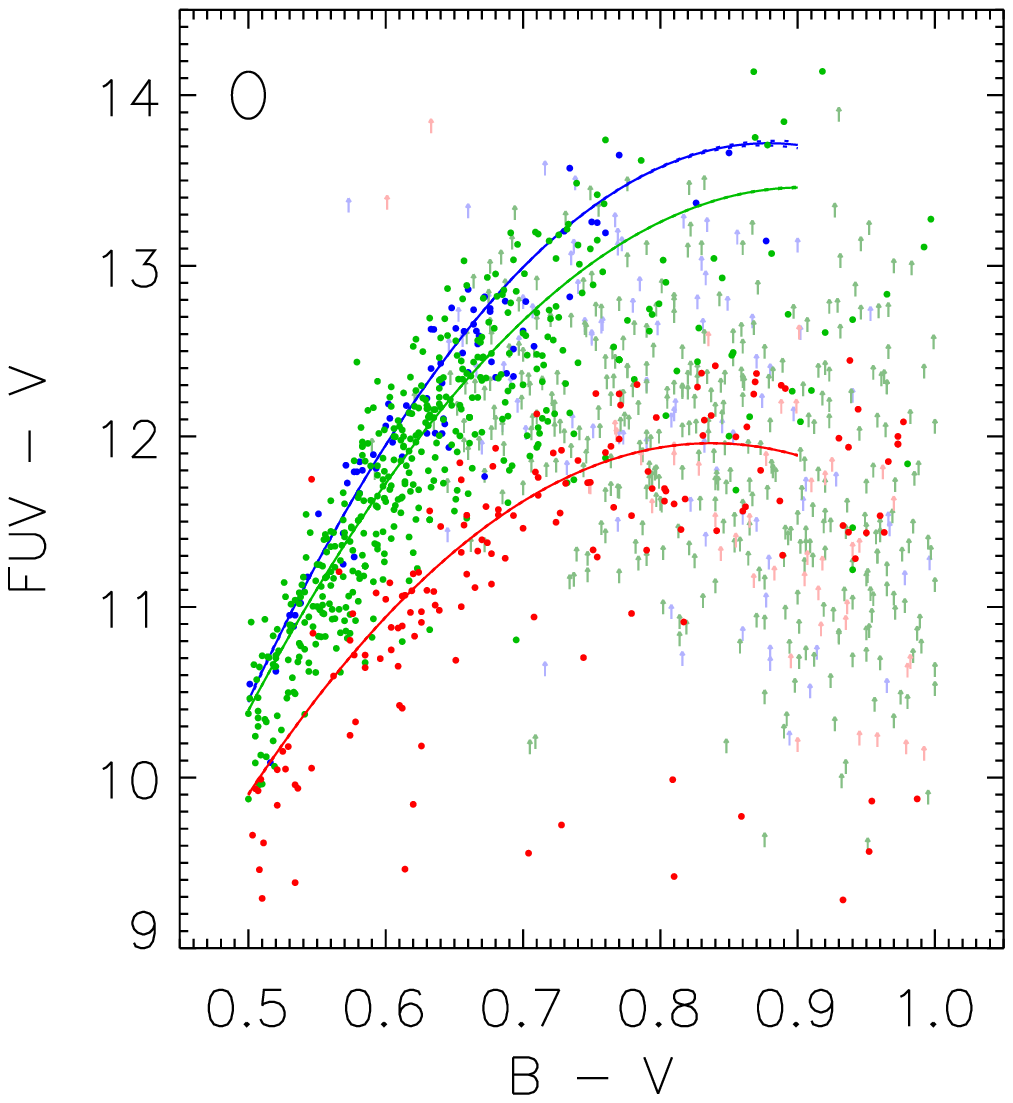}
\includegraphics[width=0.48\textwidth]{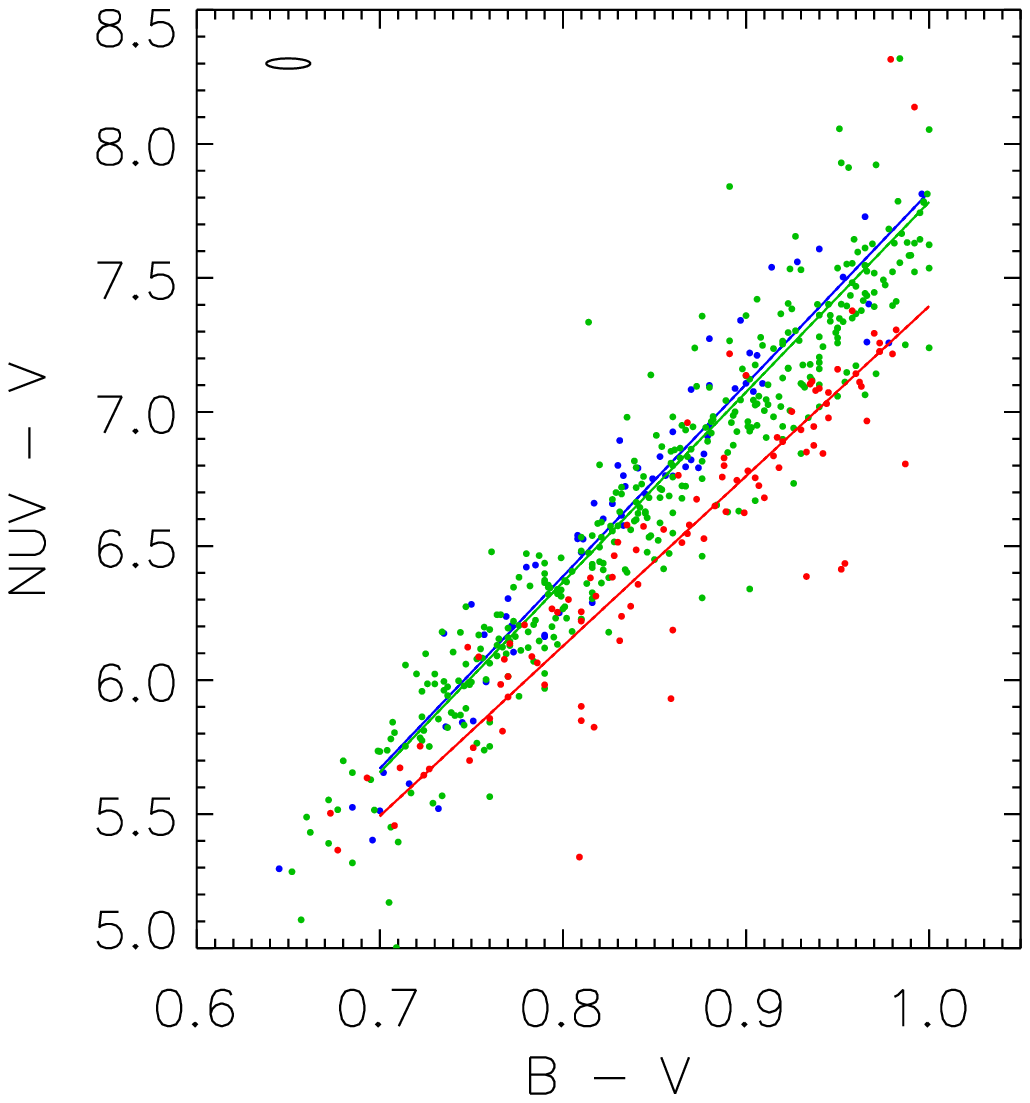}
\caption{$UV-V$ vs. $B-V$ diagrams of our volume-limited Hipparcos sample. Red points are the most active stars, with $\log{R^\prime_\textrm{HK}} > -4.5$. Green points have $-5.0 < \log{R^\prime_\textrm{HK}} < -4.5$, while blue points have $\log{R^\prime_\textrm{HK}} < -5.0$. The ellipses in the upper left corner of either panel show the median errors in color. To illustrate the dependence on activity, we plot Equations~\ref{fuvfit} and \ref{nuvfit} at the median $\log{R^\prime_\textrm{HK}}$ of each bin.}\label{ccplots_hip}
\end{figure*}

\begin{figure*}
\includegraphics[width=0.48\textwidth]{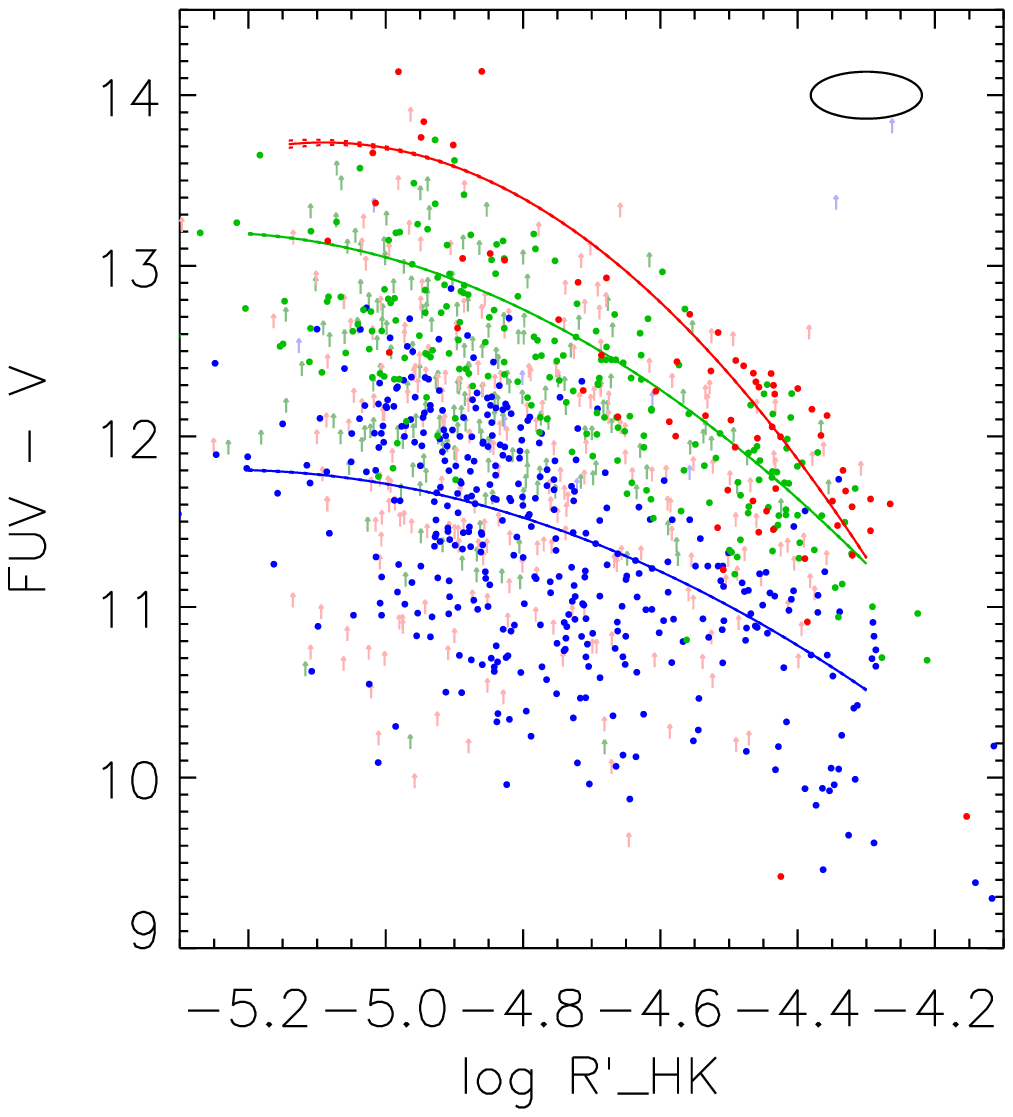}
\includegraphics[width=0.48\textwidth]{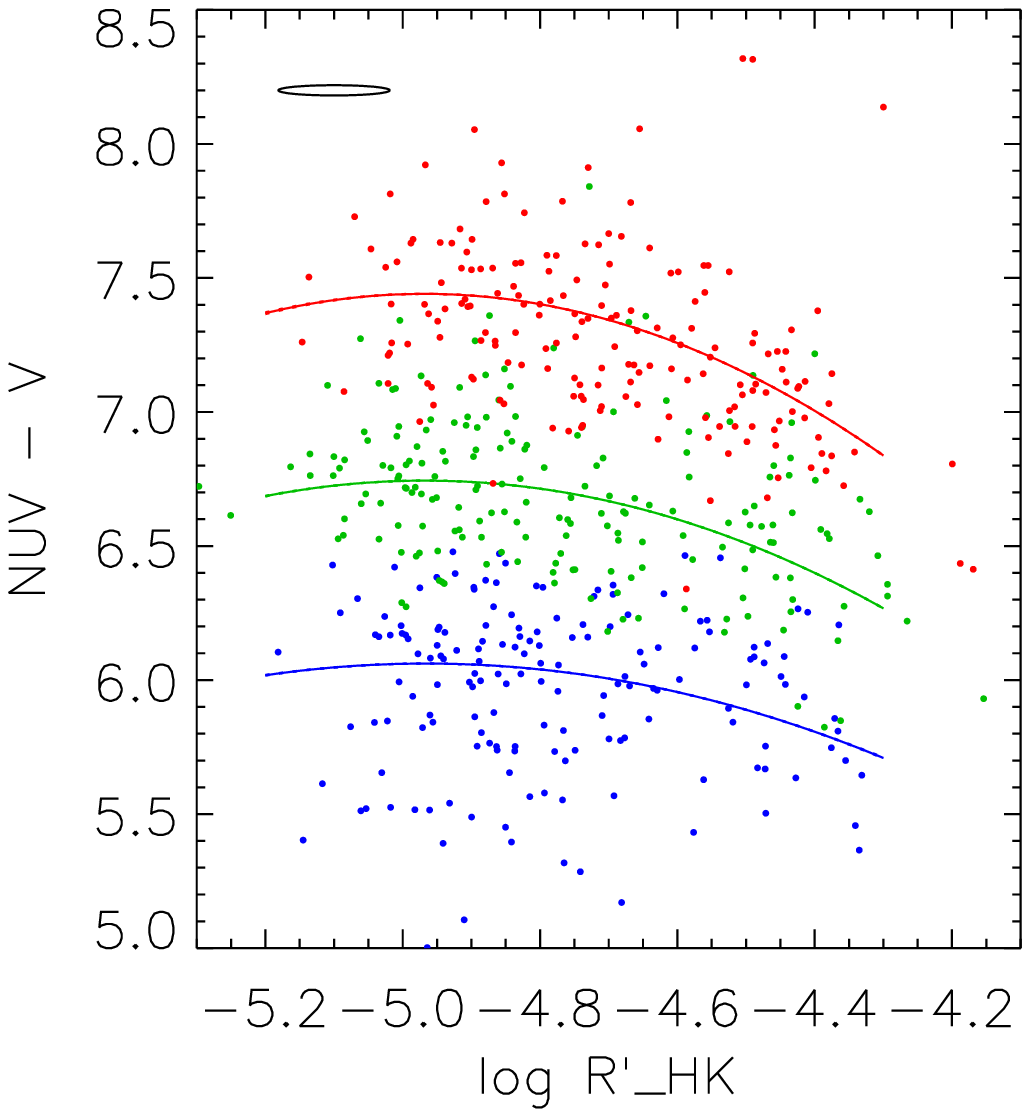}
\caption{$UV-V$ vs. $R^\prime_\textrm{HK}$ diagrams of our volume-limited Hipparcos sample. In the left panel, red points have $B-V > 0.8$, green points have $0.65 < B-V < 0.8$, and blue points have $B-V < 0.65$. In the right panel, red points have $B-V > 0.9$, green points have $0.8 < B-V < 0.9$, and blue points have $B-V < 0.8$. The ellipses at the top of either panel show the median errors in color. The utility of the UV as an activity indicator can be seen directly as a downward trend in $UV-V$ color with activity. We plot Equations~\ref{fuvfit} and \ref{nuvfit} at the median $B-V$ of each bin.}\label{uvacplots_hip}
\end{figure*}

We focused on reproducing Equation~\ref{fuvfit} and the left panel of Figure~\ref{ccplots_hip}, as this is the result that seems the most suspicious: the UV-$R^\prime_\textrm{HK}$ relation levels out at the same B-V color at which we start missing stars in the FUV, suggesting the apparent flattening is an artifact of our detection limits. We therefore constructed a model where the intrinsic UV-$R^\prime_\textrm{HK}$ relation remains linear in $B-V$ at all colors, based on a fit to stars with $B-V < 0.7$ where our observations are complete:
\def\Runit{((\log{R^\prime_\textrm{HK}})_\textrm{true}+4.5)}
\def\Bunit{((B-V)_\textrm{true}-0.8)}
\begin{eqnarray}
(FUV-V)_\textrm{true} & = & 11.19 - 2.17\Runit - 1.46\Runit^2 \nonumber \\
&&	{}+ 10.31\Bunit - 7.26\Bunit\Runit \nonumber \\
&&	{}- 8.83\Bunit\Runit^2 \label{linearmodel}
\end{eqnarray}
The linear $B-V$ dependence of this model is in contrast to Equation~\ref{fuvfit}, which has a quadratic dependence on $B-V$, leveling out over $0.7 < B-V < 0.9$.

We created a set of 1200 simulated stars by independently drawing $(V_\textrm{true}, (B-V)_\textrm{true}, (\log{R^\prime_\textrm{HK}})_\textrm{true})$ triplets from the Hipparcos sample. Use of the Hipparcos data in place of analytical models ensured our model had similar 
distributions to our data. We then applied Equation~\ref{linearmodel} to get intrinsic FUV magnitudes $FUV_\textrm{true}$. 10\% of the FUV magnitudes were brightened by an amount drawn from an exponential distribution with a mean of 1.5~mag. Introducing these bright outliers reproduced the ``cloud'' of UV-bright points seen in Figure~\ref{ccplots_hip}. Aside from the population of outliers, we assumed there was no scatter in the intrinsic UV-$R^\prime_\textrm{HK}$ relation.

To simulate the observing process, we randomly drew an exposure time and FUV sky background count rate for each star from the real GALEX observations. The exposure time and background were drawn independently from each other, and independently from the $(V, B-V, \log{R^\prime_\textrm{HK}})$ triplets, so that a typical simulated star was based on three different stars from the original data. 
Again, drawing model parameters from the data guaranteed the distributions we were using had the same properties -- such as a tail of very long exposures -- as the real data. 
We generated observed FUV fluxes for each star from a Poisson distribution with a mean equal to the intrinsic flux plus the background, multiplied by the exposure time. We calculated detection limits from Equation~\ref{uplim} and reported non-detections where the observed magnitude $FUV_\textrm{obs}$ was fainter than these limits. We generated observed $V$ magnitudes, $B-V$ colors, and $\log{R^\prime_\textrm{HK}}$ measurements simply by drawing Gaussian random numbers with means $V_\textrm{true}$, $(B-V)_\textrm{true}$, and $(\log{R^\prime_\textrm{HK}})_\textrm{true})$ and standard deviations 0.01~mag, 0.01~mag, and 0.08~dex respectively. We then fit polynomials to the data following the procedure of Section~\ref{hipresults}, including the use of the leave-one-out cross-validation score as a goodness-of-fit statistic to determine the degree of the polynomial.

The best fit by cross-validation score is linear in $B-V$ and quadratic in $\log{R^\prime_\textrm{HK}}$, like Equation~\ref{linearmodel}. The best fit that allows curvature in $B-V$ is quadratic in $B-V$ and linear in $\log{R^\prime_\textrm{HK}}$, as shown in Figure~\ref{ccplots_sim}. Figure~\ref{ccplots_sim} looks very different from Figure~\ref{ccplots_hip}, the equivalent plot in our real data. We believe the lack of curvature in the simulated fit arises because our fits to the simulated data are very poorly constrained at red $B-V$ colors. Our simulated FUV observations detect only 37 of 394 stars (9\%) with $B-V \ge 0.8$, and 6 of 177 stars (3\%) with $B-V \ge 0.9$, compared to our real detection rate of 81 of 396 (20\%) with $B-V \ge 0.8$ and 29 of 184 (16\%) with $B-V \ge 0.9$. This is because Equation~\ref{linearmodel} places most cool stars 2 magnitudes below our sensitivity limits, so only the ``outliers'' -- the 10\% of stars that were artificially given an FUV excess -- are detected.
To get the simulations to reproduce the observed detection rate at red $B-V$, we need to artificially brighten $\sim 60\%$ of the FUV magnitudes rather than 10\%. Not only does brightening so many stars increase the number of UV-bright outliers at blue $B-V$ colors well beyond the observed number, it means that Equation~\ref{linearmodel} no longer describes the simulated population as intended.

We conclude that the intrinsic $FUV-V$ vs. $B-V$ relation does flatten at $B-V \gtrsim 0.8$, as otherwise we would expect fewer detections and a qualitatively different fit than we observe. However, since Figure~\ref{ccplots_sim} shows the fit to the simulated data is shallower than Equation~\ref{linearmodel}, it is still possible that Equation~\ref{fuvfit} exaggerates the flatness of the intrinsic $FUV-V$ vs. $B-V$ relation.

\subsection{Testing the Dependence on Functional Form}\label{nonparatest}

In Sections~\ref{hipresults} and \ref{mgresults}, we fit our data with bivariate polynomials, choosing the degree of the polynomial that minimized the leave-one-out cross validation score $CV$ \citep[e.g.][]{crossvalidation}, which acted as a goodness-of-fit statistic.
However, the solution with the lowest $CV$ score was often one of several with very similar goodness of fit. Since $CV$, because of its generality, does not follow a specific distribution such as the chi-squared distribution, we cannot judge whether a difference in $CV$ is statistically significant. This potentially makes the choice of the best form for the fit sensitive to fluctuations in $CV$. 

Fortunately, the functional form of the fit -- in our case, the degree of the polynomial -- carries no physical significance. The decision to choose one form over another is only important insofar as it changes the value of $UV-V$, $R^\prime_{HK}$, $NUV-J$, or age predicted for a particular star. In this section, we show that replacing the polynomial fit with the lowest $CV$ score with, for example, the fit that gives the second-lowest score does not significantly change the numerical values predicted by the fit. It follows that fluctuations in $CV$ do not matter: all the fits with similar $CV$ scores make the same predictions.

For each of Equations~\ref{fuvfit}-\ref{nacfit} and \ref{nuvjfit}-\ref{nacjfit}, we compared the fits we presented in Sections~\ref{hipresults} and \ref{mgresults}, which had the lowest $CV$ score, with the alternative polynomials that had the second, third, and fourth lowest scores. 
For example, our Equation~\ref{fuvfit} presents $FUV-V$ as a quadratic function of $B-V$ and a quadratic function of $\log{R^\prime_\textrm{HK}}$. The model that had the second-lowest $CV$ score gave $FUV-V$ as a cubic function of $B-V$ and a quadratic function of $\log{R^\prime_\textrm{HK}}$, but with best-fit parameters that caused the two models to predict the same value of $FUV-V$, to within 0.03~mag, over most of the $B-V$ vs. $\log{R^\prime_\textrm{HK}}$ parameter space. It follows that the choice of polynomial form for Equation~\ref{fuvfit}, quadratic or cubic in $B-V$, matters only at the $\sim 0.03$~mag level.

In Table~\ref{deltaf}, we present the RMS difference between each model adopted in Equations~\ref{fuvfit}-\ref{nacfit} and \ref{nuvjfit}-\ref{nacjfit} and each of its three alternatives, with the RMS taken over the region in parameter space where each model is valid. For comparison, we also list the residuals we found around the equations.
The RMS differences between alternate forms for the fits are typically a factor of 3-4 less than the scatter around the fits, indicating that choosing one of the alternate forms instead would not have changed our results.

\subsection{Testing for Selection Effects}\label{samplebias}

\label{biastests}
\begin{figure*}
\includegraphics[width=0.48\textwidth]{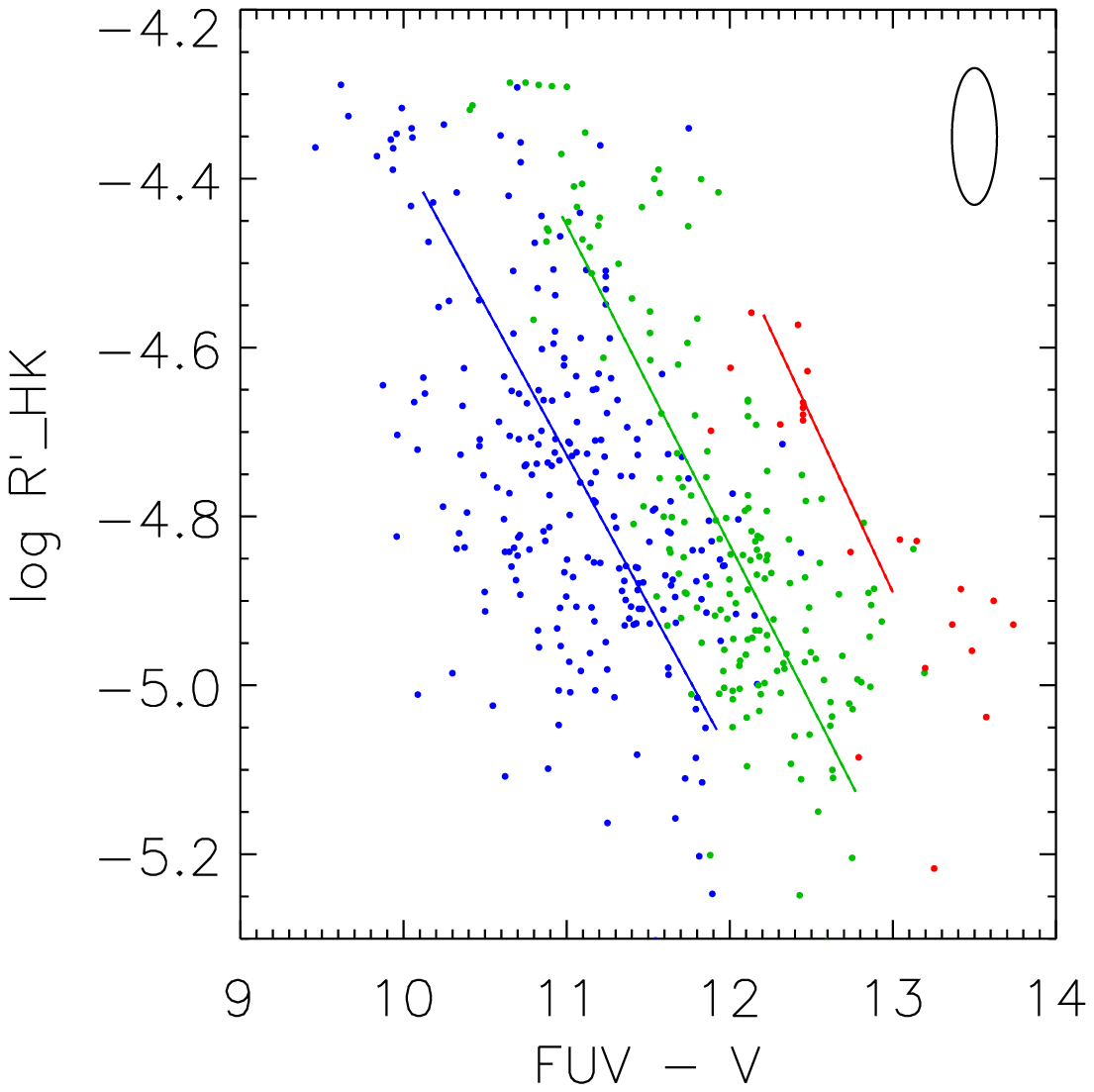}
\includegraphics[width=0.48\textwidth]{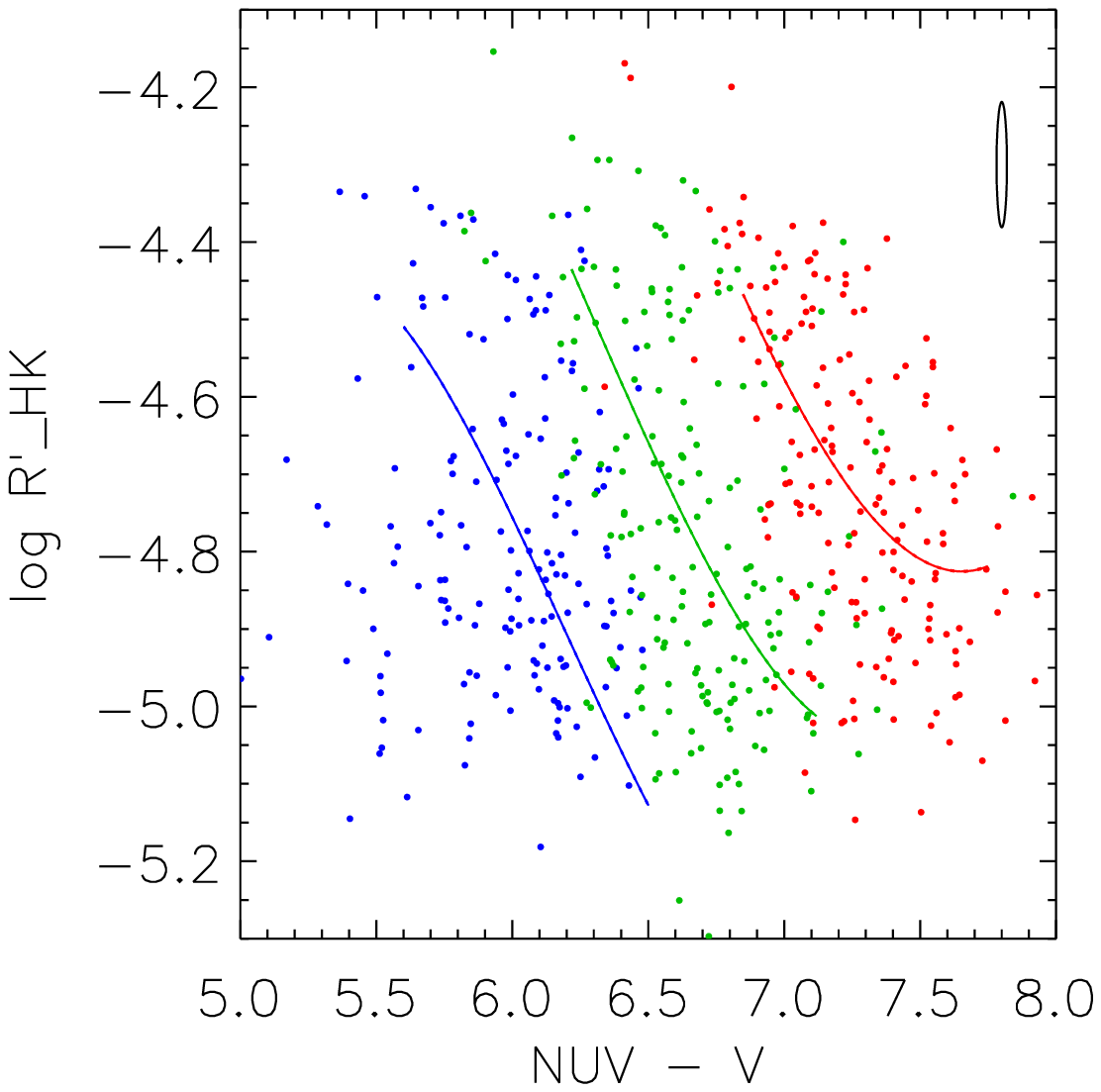}
\caption{$R^\prime_\textrm{HK}$ vs. $UV-V$ diagrams of our volume-limited Hipparcos sample. The left panel only shows stars with $V + 12.03 (B-V) < 15.5$, where our FUV observations were complete.
In the left panel, red points have $B-V > 0.7$, green points have $0.6 < B-V < 0.7$, and blue points have $B-V < 0.6$. In the right panel, red points have $B-V > 0.9$, green points have $0.8 < B-V < 0.9$, and blue points have $B-V < 0.8$. The ellipses at the top right of either panel show the median errors along each axis. The main difference between these figures and Figure~\ref{uvacplots_hip} is that we plot Equations~\ref{facfit} and \ref{nacfit}, which give the average $R^\prime_\textrm{HK}$ at fixed $UV-V$ rather than the average $UV-V$ at fixed $R^\prime_\textrm{HK}$.}\label{acuvplots_hip}
\end{figure*}

\begin{figure*}
\includegraphics[width=0.48\textwidth]{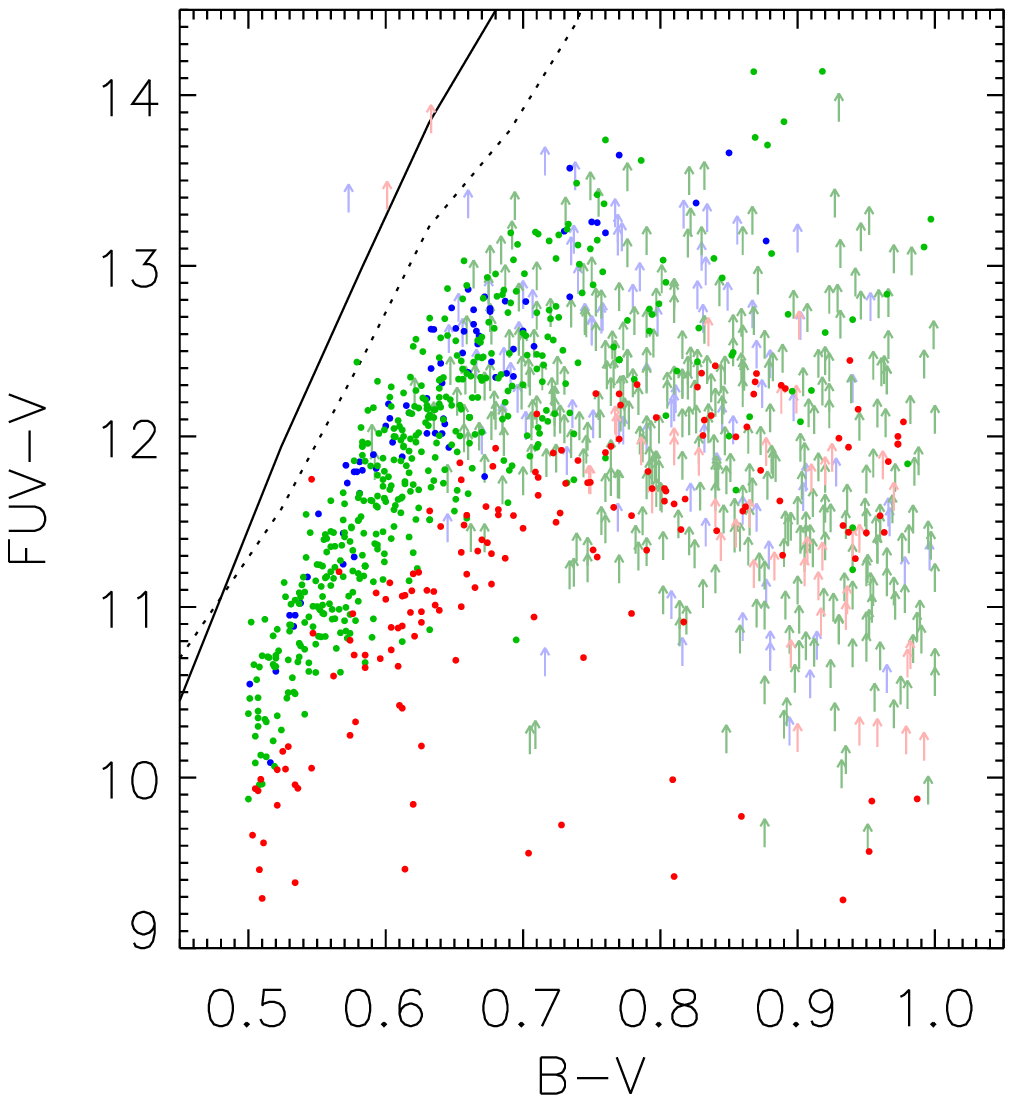}
\includegraphics[width=0.48\textwidth]{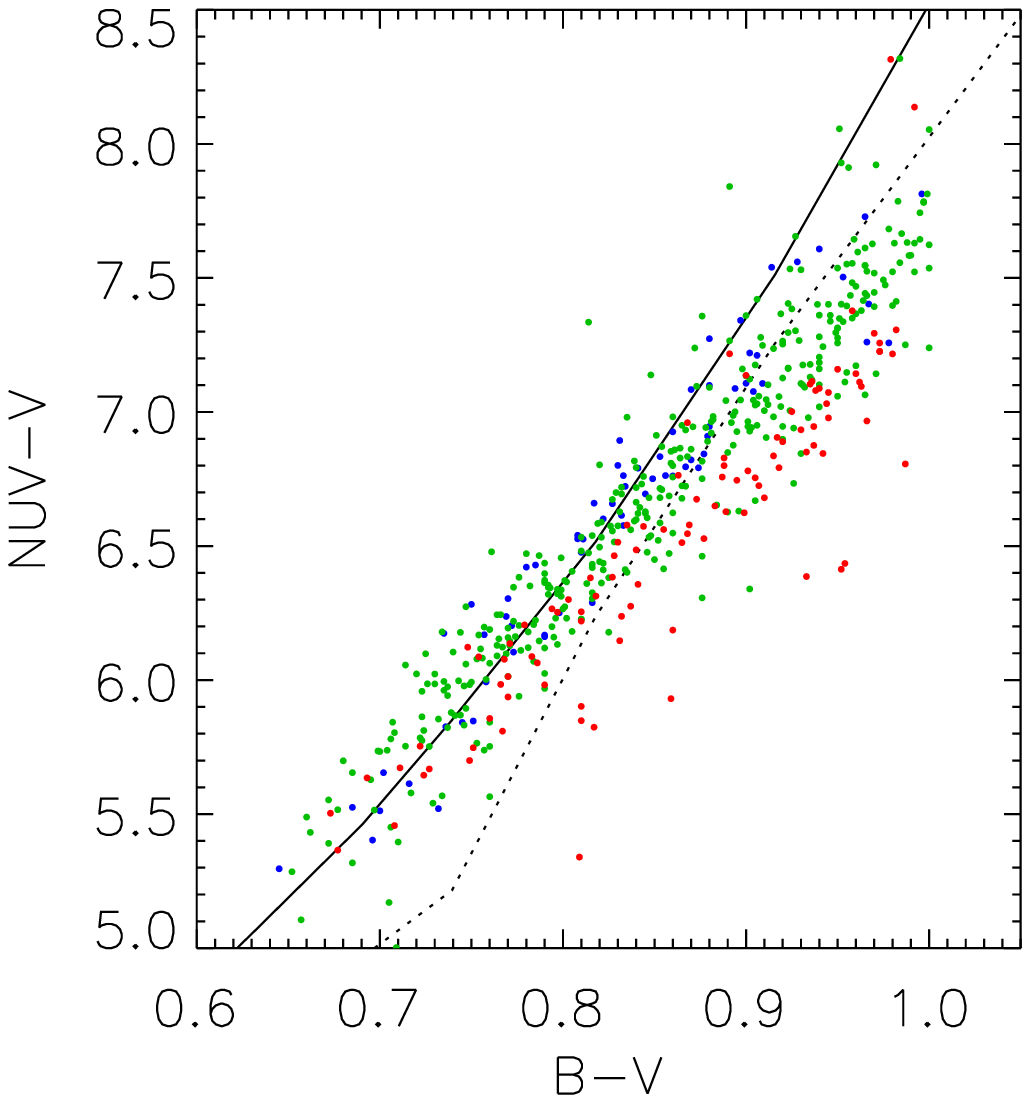}
\caption{Color-color diagrams comparing our synthetic UV fluxes to our observations with GALEX. Colors have the same meaning as in Figure~\ref{ccplots_hip}. The dotted curve shows the stellar locus predicted by matching the synthetic FUV and NUV magnitudes calculated for each model temperature with the photometry listed by \citet{mscolors} for the same temperature. The results differ markedly from the observations, particularly in the NUV, where the synthetic magnitudes imply that a large fraction of stars emit \emph{less} UV flux than their photospheres can account for. The solid curve shows the stellar locus after the photosphere FUV and NUV magnitudes have been corrected for temperature systematics as described in the text; the results appear much more consistent with the NUV observations.}\label{synthuvmags}
\end{figure*}

\begin{figure*}
\includegraphics[width=0.48\textwidth]{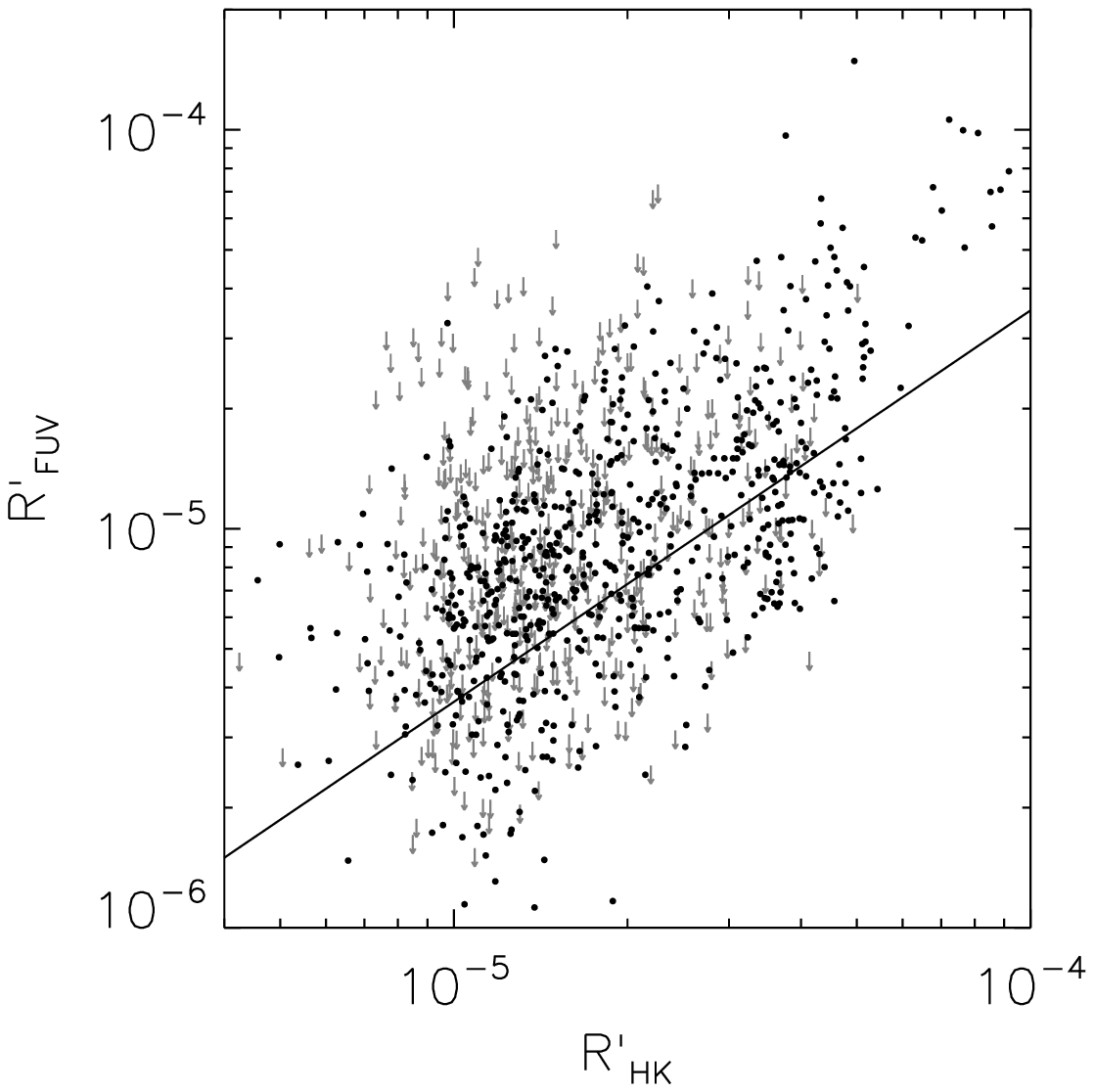}
\includegraphics[width=0.48\textwidth]{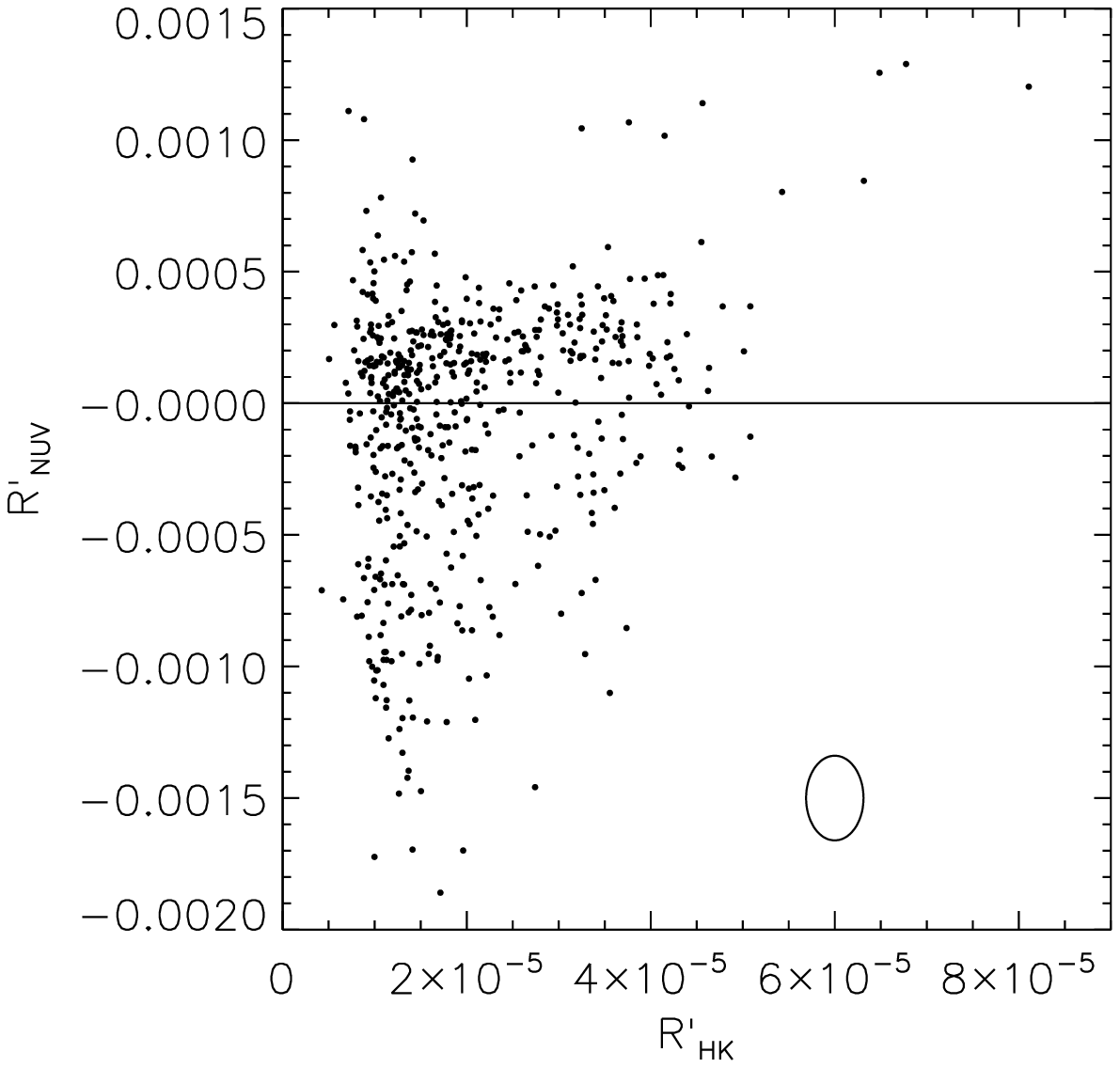}
\caption{$R^\prime_\textrm{UV}$ vs. $R^\prime_\textrm{HK}$ plots of our volume-limited Hipparcos sample. The left panel, a log plot, shows the correlation between $R^\prime_\textrm{FUV}$ and $R^\prime_\textrm{HK}$. The solid line represents the fit given by Equation~\ref{hip_ruvfit}. The right panel, a linear plot, shows that $R^\prime_\textrm{NUV}$ measurements are very noisy with many negative values. The ellipse in both panels shows the median errors. The distribution of $R^\prime_\textrm{NUV}$ is centered on zero.}\label{activity2}
\end{figure*}

\begin{figure*}
\includegraphics[width=0.48\textwidth]{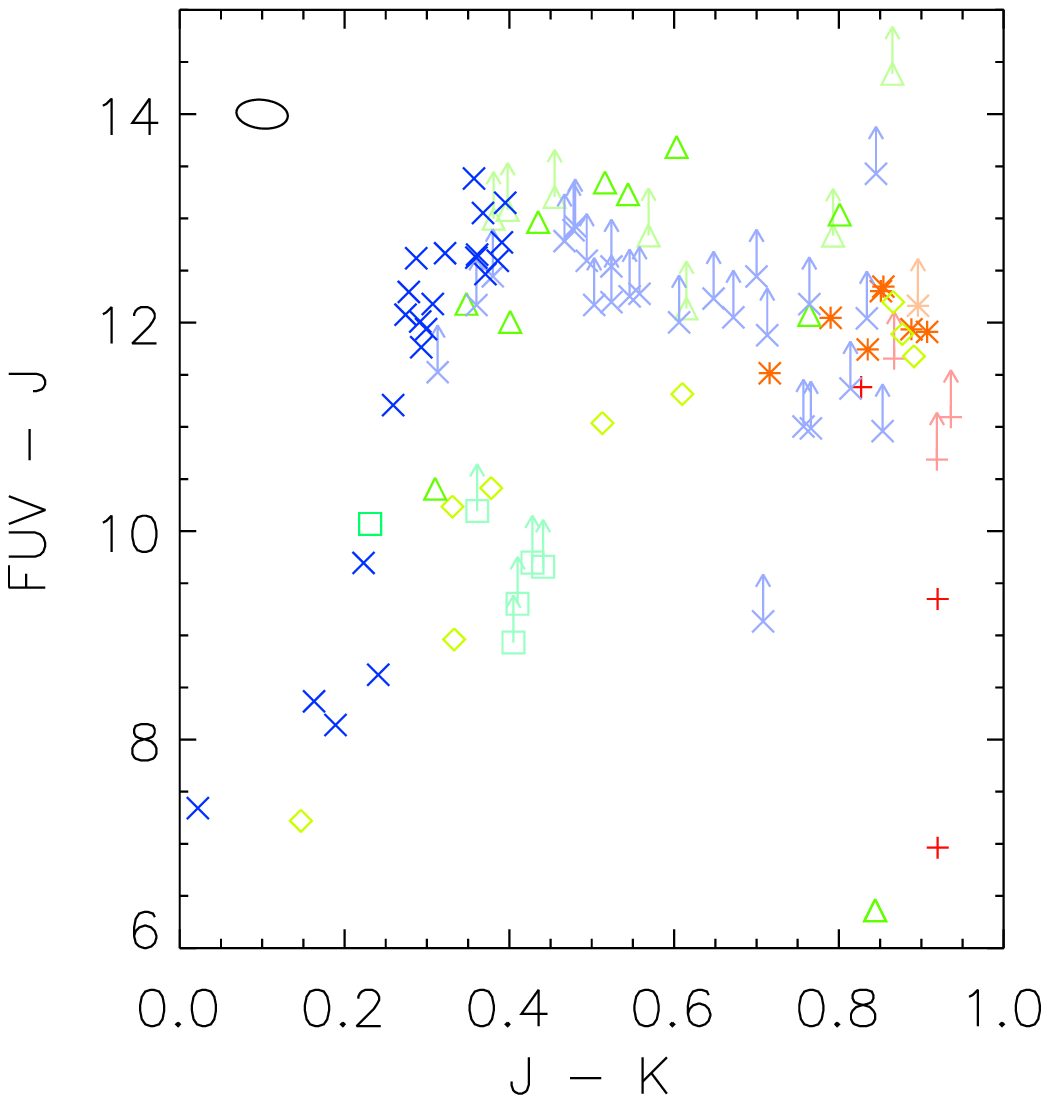}
\includegraphics[width=0.48\textwidth]{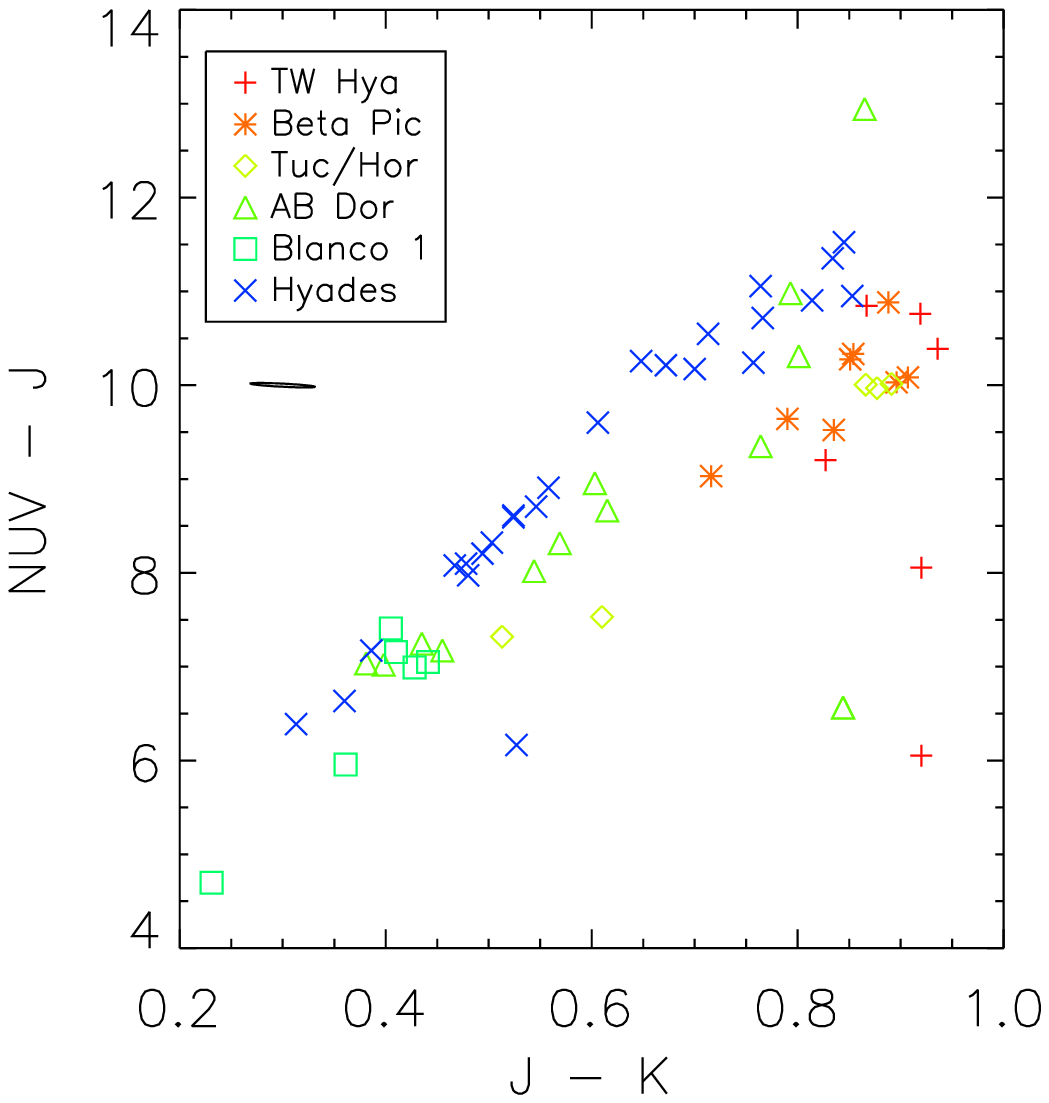}
\caption{$UV-J$ vs. $J-K$ diagrams of our cluster and moving group sample. Each symbol corresponds to a different association, as shown in the legend, while the hue maps linearly to $\log{\textrm{age}}$. The ellipses at the upper left of either figure show the median errors in color. The NUV panel shows a clear trend with age, but any relation between FUV and age is obscured by non-detections.}\label{ccplots_mg}
\end{figure*}

\begin{figure*}
\includegraphics[width=0.48\textwidth]{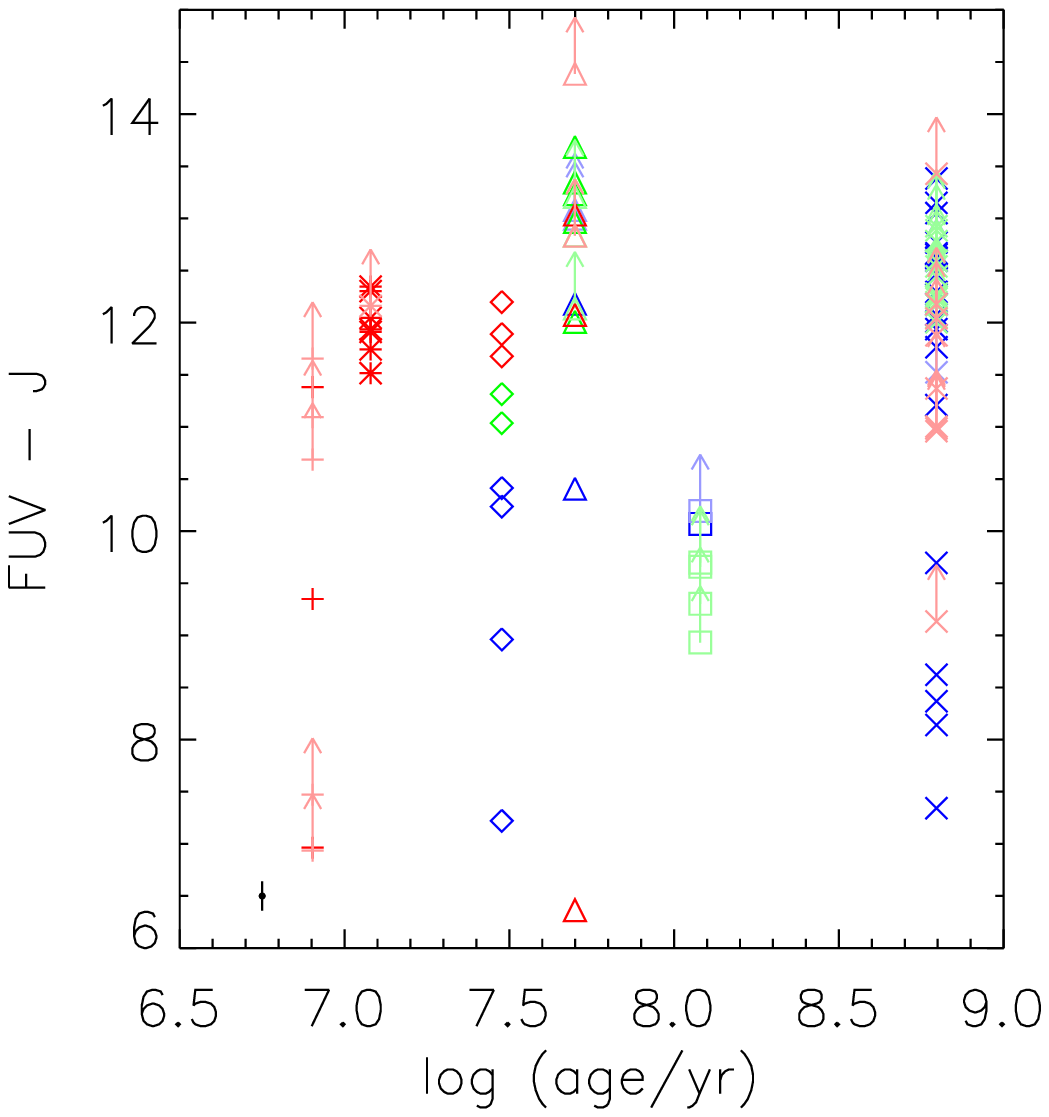}
\includegraphics[width=0.48\textwidth]{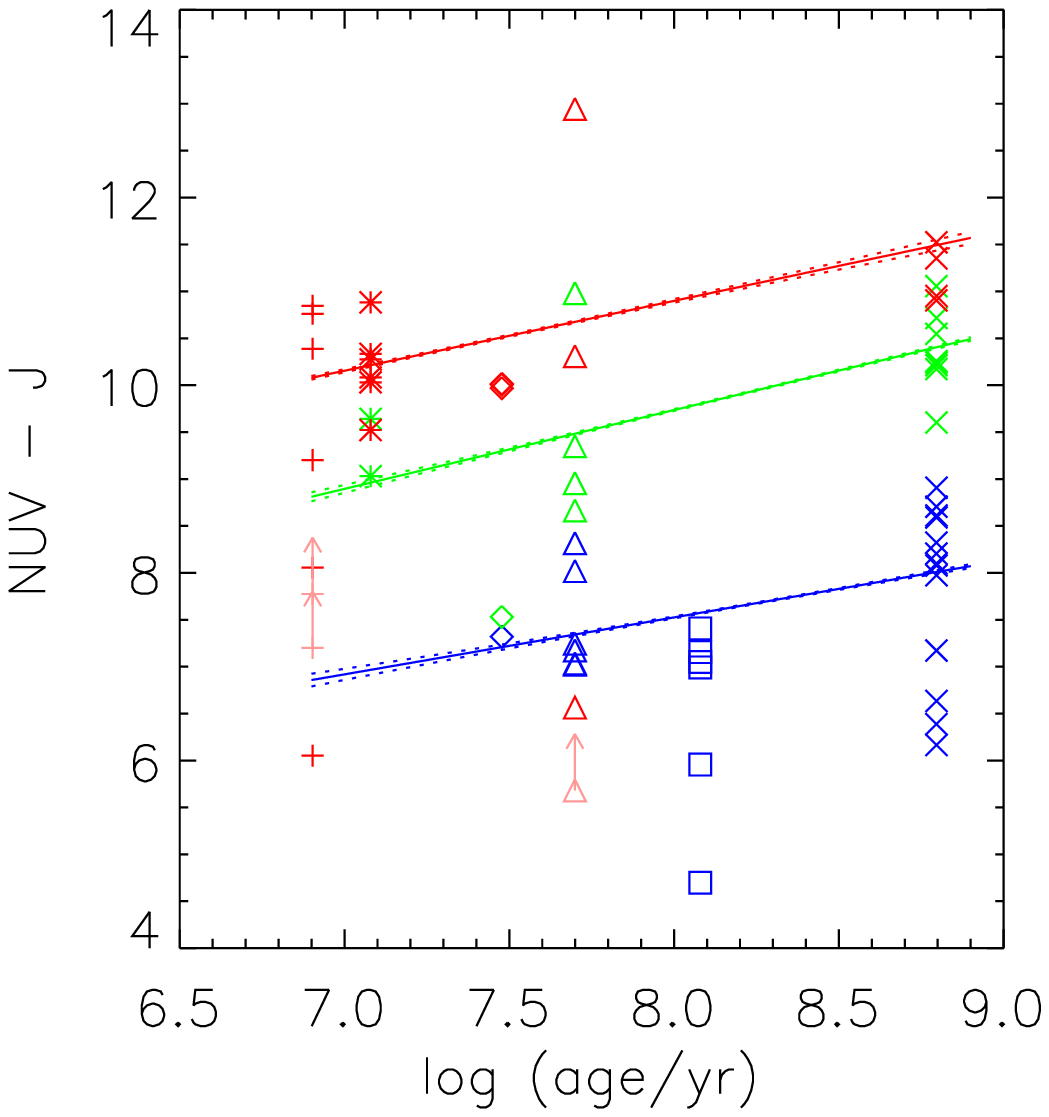}
\caption{$UV-J$ vs. age diagrams  of our cluster and moving group sample. Symbols are the same as in Figure~\ref{ccplots_mg}. In the left panel, red points have $J-K > 0.7$, green points have $0.4 < J-K < 0.7$, and blue points have $J-K < 0.4$. In the right panel, red points have $J-K > 0.8$, green points have $0.6 < J-K < 0.8$, and blue points have $J-K < 0.6$. To illustrate the slight trend with age, we plot Equation~\ref{nuvjfit} at the median $J-K$ of each bin.}\label{uvacplots_mg}
\end{figure*}

\begin{figure*}
\includegraphics[width=0.48\textwidth]{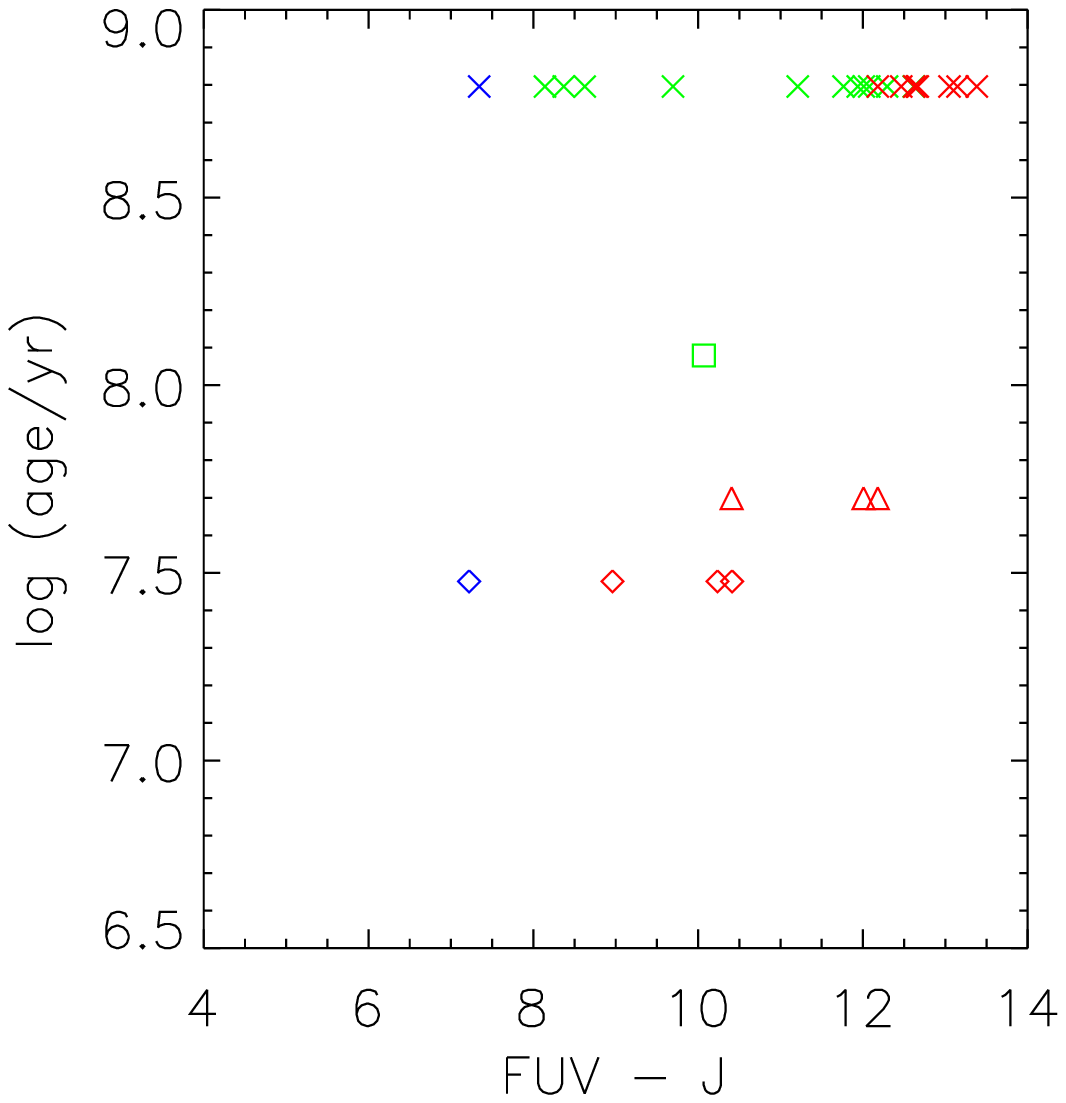}
\includegraphics[width=0.48\textwidth]{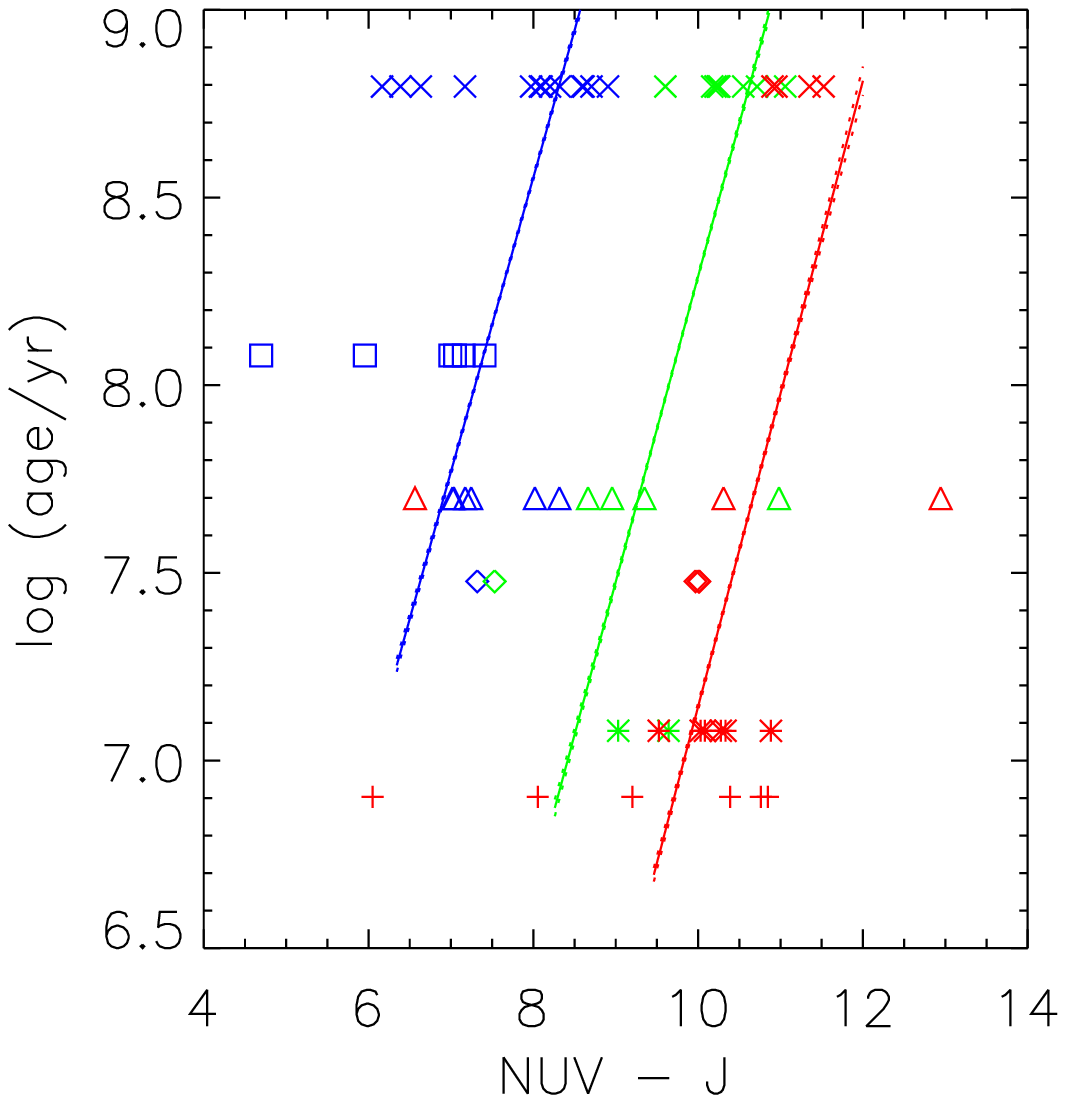}
\caption{Age vs. $UV-J$ diagrams of our cluster and moving group sample. Symbols are the same as in Figure~\ref{ccplots_mg}. 
The left panel only shows stars with $J + 17.29 (J-K) < 13.87$, where our FUV observations were complete.
In the left panel, red points have $J-K > 0.3$, green points have $0.2 < J-K < 0.3$, and blue points have $J-K < 0.2$. In the right panel, red points have $J-K > 0.8$, green points have $0.6 < J-K < 0.8$, and blue points have $J-K < 0.6$. The main difference between these figures and Figure~\ref{uvacplots_mg} is that we plot Equation~\ref{nacjfit}, which gives the average age at fixed $NUV-J$ rather than the average $NUV-J$ at fixed age. While UV-bright stars do tend to be older than UV-faint stars on average, there is so much scatter at fixed age that the age of any individual star cannot be predicted well.}\label{acuvplots_mg}
\end{figure*}

\begin{figure*}
\includegraphics[width=0.48\textwidth]{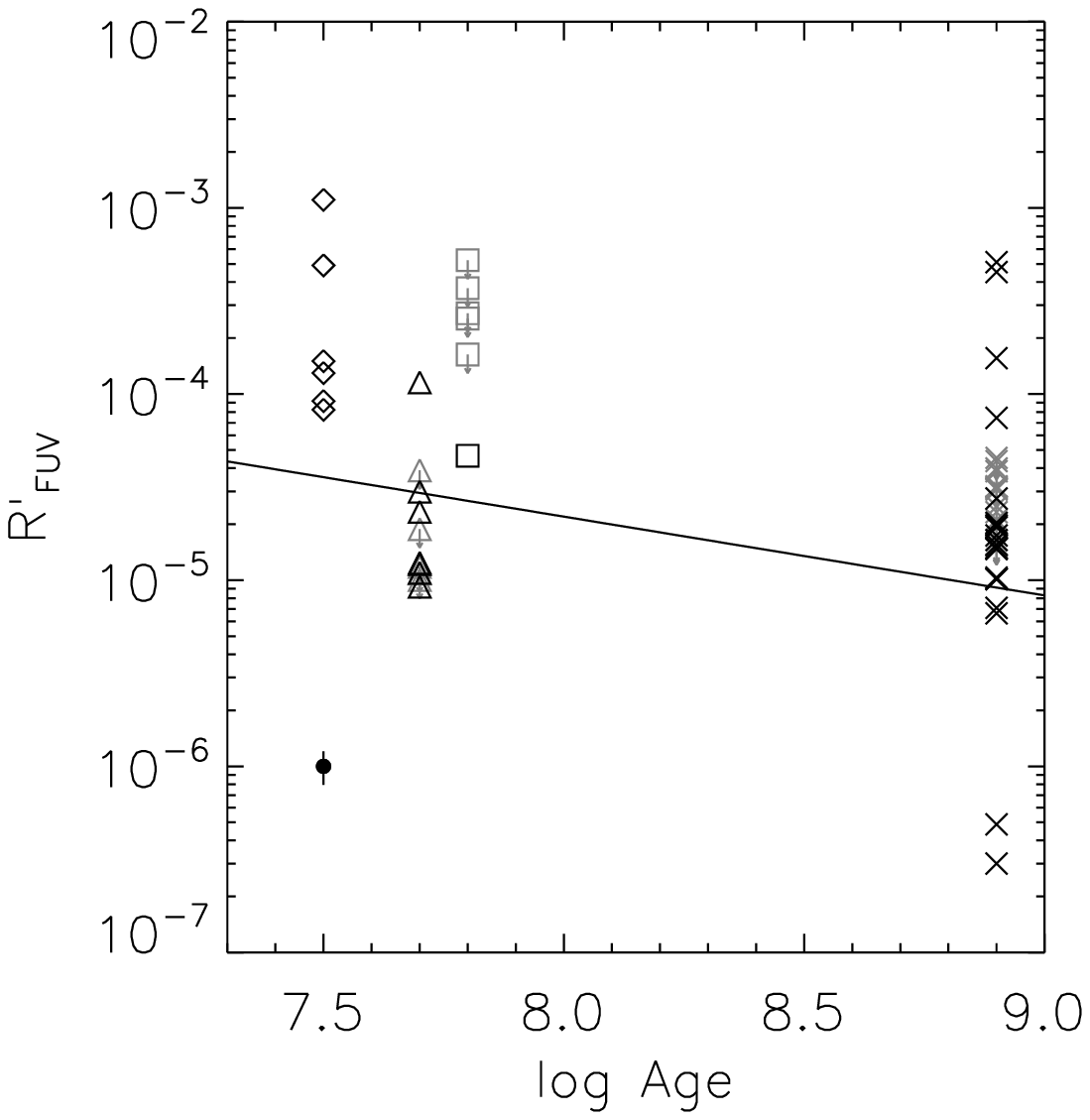}
\includegraphics[width=0.48\textwidth]{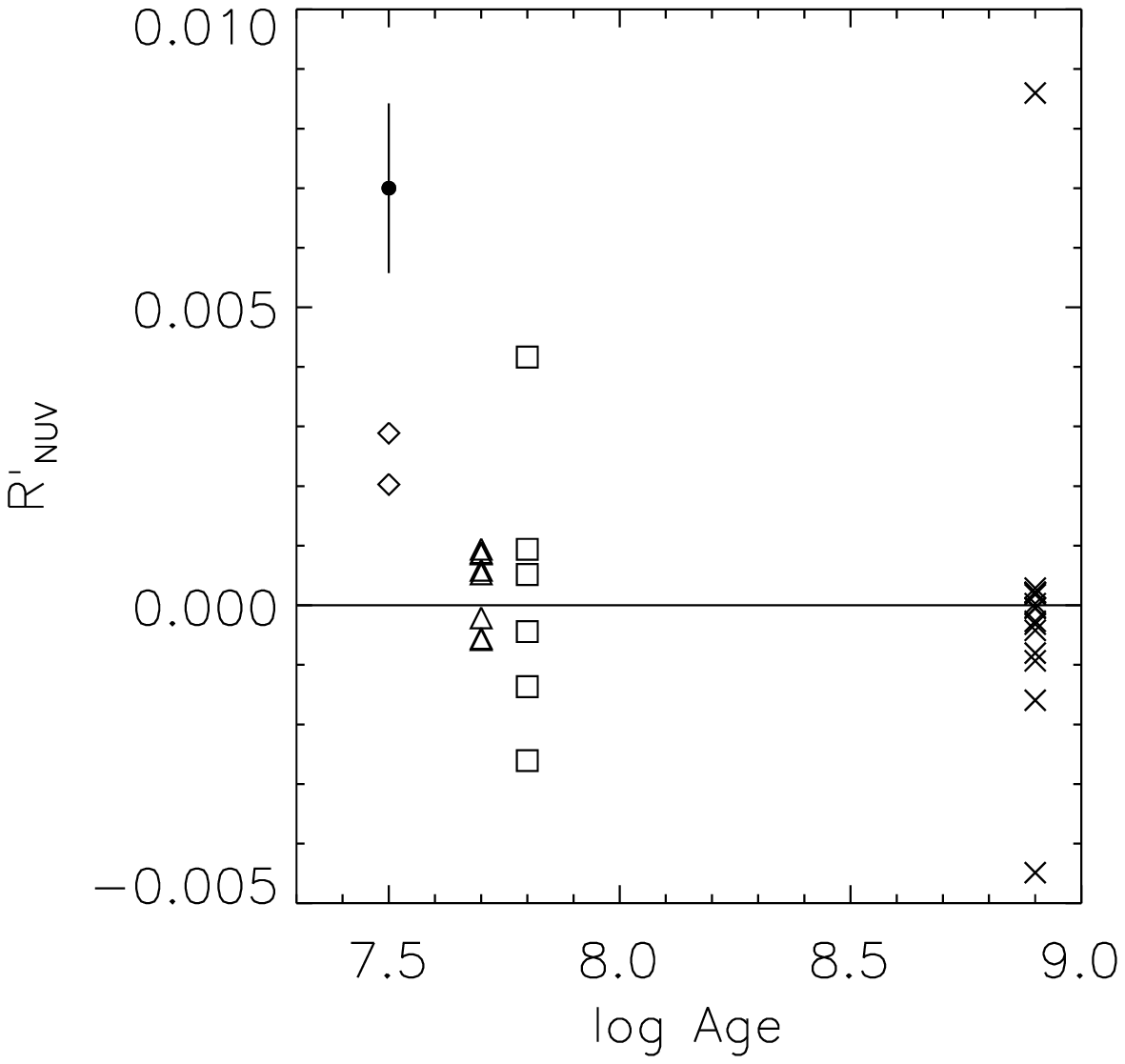}
\caption{$R^\prime_\textrm{UV}$ vs. age plots of our moving group and cluster sample. Symbols are the same as in Figure~\ref{ccplots_mg}. The left panel, a log plot, shows the correlation between $R^\prime_\textrm{FUV}$ and age. The solid line represents the fit given by Equation~\ref{mg_ruvfit}. The right panel, a linear plot, shows that $R^\prime_\textrm{NUV}$ measurements appear to be dominated by noise. The filled circle in either panel is not a real point, but illustrates the median error in $R^\prime_\textrm{UV}$. Both panels include only stars whose inferred spectral type is K5 or hotter, as we lack the models to estimate photospheric UV fluxes for cooler stars. Thus, there are fewer points than in Figure~\ref{uvacplots_mg}, in particular none younger than $\sim 30$~Myr.}\label{activityage}
\end{figure*}

\begin{figure*}
\includegraphics[width=0.48\textwidth]{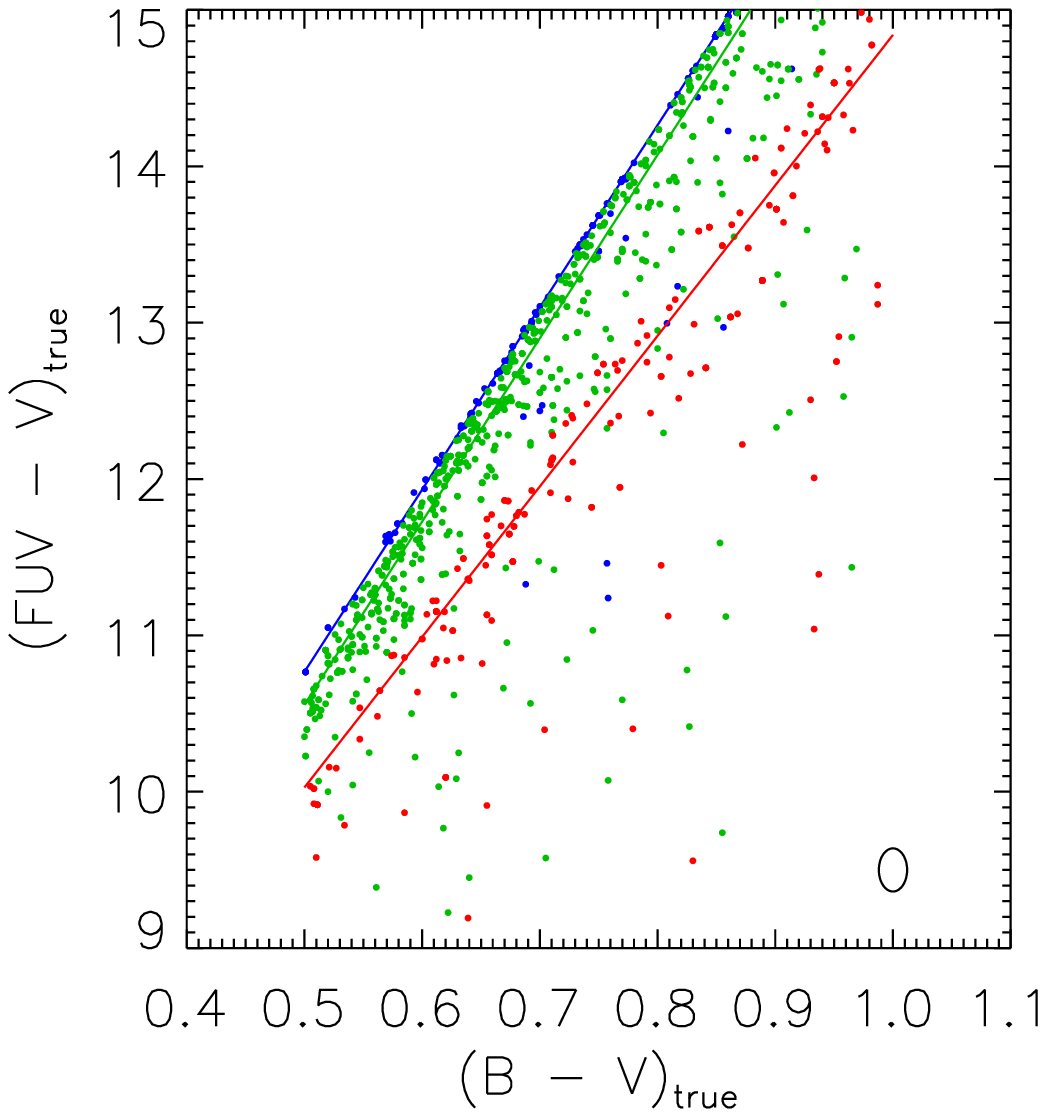}
\includegraphics[width=0.48\textwidth]{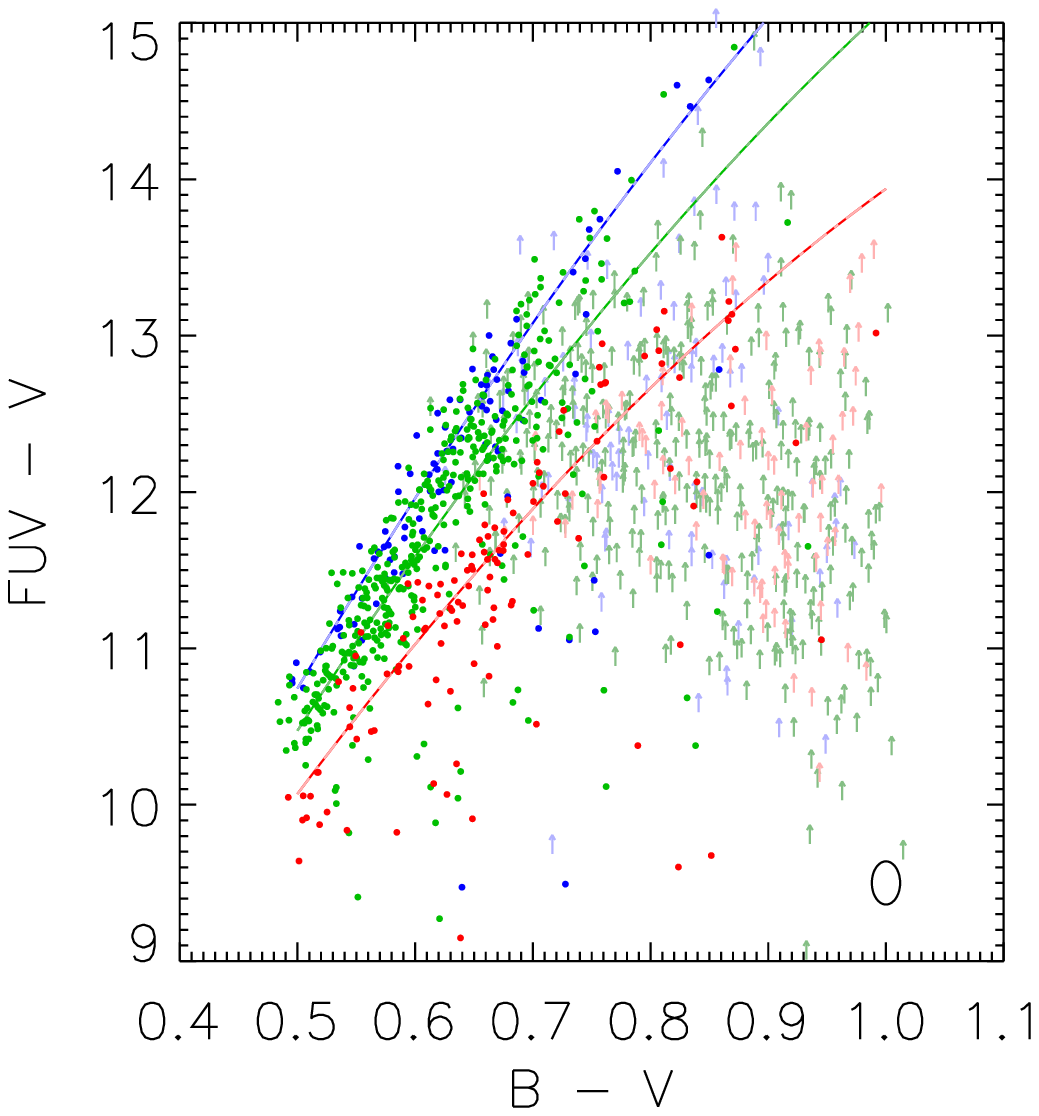}
\caption{A typical run of our Monte Carlo simulations in Section~\ref{montecarlo}. The panel on the left shows the intrinsic color-color relation, without observational errors or selection effects, together with Equation~\ref{linearmodel}, which we used to generate the data. The panel on the right shows the same sample, but with observational errors and flux limits applied, and a fit to the simulated data.
On both panels red points are the most active stars, with $\log{R^\prime_\textrm{HK}} > -4.5$, green points have $-5.0 < \log{R^\prime_\textrm{HK}} < -4.5$, while blue points have $\log{R^\prime_\textrm{HK}} < -5.0$. Equation~\ref{linearmodel} and our fit to the simulated data are plotted at the median $\log{R^\prime_\textrm{HK}}$ of each bin.
}\label{ccplots_sim}
\end{figure*}

The trend between UV color and $R^\prime_\textrm{HK}$ seen in Figures~\ref{ccplots_hip}-\ref{acuvplots_hip} is small, comparable to the scatter in the data. Here we show that this trend is not the result of flux biases, such as active stars being more distant or less luminous and therefore having UV fluxes more prone to measurement error.

We binned the data in $\log{R^\prime_\textrm{HK}}$ and $B-V$ and calculated the average apparent $V$ magnitude and average parallax for each bin. We present the means and the errors on the means in Table~\ref{biastable}. In each $B-V$ range, we used a two-sided t test to test whether the means in different $\log{R^\prime_\textrm{HK}}$ bins were significantly different. 
No pair of mean parallaxes differs at more than 82\% confidence. For stars with $B-V > 0.7$, only one pair of mean magnitudes differs at more than 95\% confidence, which is not a significant result in 9 tests. However, for stars with $B-V < 0.7$ the least active bin ($\log{R^\prime_\textrm{HK}} < -5.0$) differs from either of the other two bins at 99.9\% confidence or greater. Inactive stars with $0.5 < B-V < 0.6$ are on average 0.5 magnitudes brighter than their active counterparts, while those with $0.6 < B-V < 0.7$ are 0.3 magnitudes brighter. This correlation represents the fact that evolved main sequence stars tend to be both less active and more luminous than their ZAMS counterparts.

Since the only correlation between apparent magnitude and $R^\prime_\textrm{HK}$ is at the blue end of our sample, whereas our inferred relationships between UV flux and $R^\prime_\textrm{HK}$ are strongest for the reddest stars in our sample, we conclude our results are not caused by a bias in stellar magnitude with $R^\prime_\textrm{HK}$.

\subsection{Testing for the Effects of Source Color}\label{colortests}

In Section~\ref{rprimeuv}, we inferred excess UV fluxes from our GALEX photometry. The inferred fluxes are proportional to the instrument count rate $CPS = \int{A_\textrm{eff}(\nu) f_\nu(\nu)/(h\nu) d\nu}$, where $A_\textrm{eff}$ is the effective area for the FUV or NUV detector. However, we interpret the fluxes as the amount of emission in a well-defined band, $f_\textrm{UV} = \int_{\nu_1}^{\nu_2}{f_\nu(\nu) d\nu}$, which does not correlate perfectly with count rate. Here we show that the color of the source does not introduce large inaccuracies in our inferred $f_\textrm{UV}$.

We considered three templates for $f_\nu$: a $10^4$~K blackbody, a $10^5$~K blackbody, and a list of FUV lines in the $\alpha$~Cen spectrum of \citet{alphacen}. The first two models bracket the temperature range at which most FUV and NUV emission lines and continuum form. The third allows us to consider a third extreme, a spectrum dominated entirely by lines with no continuum contribution. Stellar UV spectra are a superposition of $10^{4-5}$~K emission from both lines and continuum, so analyzing the color corrections for the three models places an upper limit on the variation we expect to see in real data.

Evaluating the integrals numerically, we normalized each template spectrum to give the same CPS in the FUV or NUV band, as appropriate. We then evaluated $f_\textrm{UV}$ directly from the scaled spectrum.
We found that the $10^5$~K blackbody that produces the same FUV GALEX response as the $10^4$~K blackbody has 92\% the latter's flux in the 1350-1780\AA\ range. The $\alpha$~Cen line list has 97\% the flux of the $10^4$~K blackbody. The $10^5$~K blackbody that produces the same NUV response as the $10^4$~K blackbody has 107\% the flux in 1780-2830\AA. We could not evaluate the NUV response to the $\alpha$~Cen line list because it only extends to 1690\AA.

In summary, the true flux associated with a GALEX magnitude varies by up to 8\% over the range of templates we have adopted.
Since the three template spectra bracket the variety of source colors we expect to encounter, we likewise expect that color systematics will not cause the true flux to deviate by more than 8\% from our estimate.

\section{Discussion}\label{discuss}

%

\subsection{UV Flux Evolution with Age}

\label{rprimecompare}

In our data, both GALEX bands probe multiple physical environments. The NUV flux is dominated by the photosphere, as seen in the close agreement of synthetic and observed NUV colors in Figure~\ref{synthuvmags}. However, the stellar chromosphere, i.e. optically thin gas cooler than $\sim 1-2 \times 10^4$~K, also contributes to the NUV flux through lines from a variety of low-ionization species. 
The FUV flux is, to order of magnitude, evenly split (see Appendix~\ref{fuvfrac}) between continuum emission from the chromosphere and lines from both the chromosphere and the transition region (a narrow zone of $10^4$~K to $\sim 5 \times 10^5$~K gas). Our results on stellar activity in the UV need to be interpreted with these mixtures in mind.

The most active stars in our Hipparcos sample typically show $R^\prime_\textrm{FUV} \sim 10^{-4.5}$; those in the moving group and cluster sample can reach up to $R^\prime_\textrm{FUV} \sim 10^{-3.5}$.
\citet{diskevol1} found, by integrating over spectra from \citet{valenti}, that weak-lined T~Tauri stars show a median $R^\prime_\textrm{913-2070\AA} \sim 10^{-3.3}$. Their figure qualitatively agrees with our results, if we assume that 50-80\% of the excess flux in 913-2070\AA\ is emitted by the Lyman $\alpha$ line, and if we also note that our FUV band (1350-1780\AA) is about one third as wide as theirs.

Our relations between the FUV excess flux and the Ca~II H and K line flux are likewise consistent with previous work on correlations between the Ca~II and UV lines. The most extensive such comparison was by \citet{activityvsactivity}, who found:
\begin{eqnarray*}
\log{F_\textrm{Si~II}} & = & 1.36 \log{F_\textrm{HK}} - 3.8 \\
\log{F_\textrm{C~II}}  & = & 1.65 \log{F_\textrm{HK}} - 6.2 \\
\log{F_\textrm{Si~IV}} & = & 2.14 \log{F_\textrm{HK}} - 9.3 \\
\log{F_\textrm{C~IV}}  & = & 1.85 \log{F_\textrm{HK}} - 7.2
\end{eqnarray*}
where $F$ denotes the flux in a particular line at the stellar surface. \citeauthor{activityvsactivity} did not subtract a contribution from the photosphere when calculating their Ca~II H and K surface fluxes, so to compare to their results we also avoid subtracting a photosphere contribution. Following the procedure we used to fit Equation~\ref{hip_ruvfit}, but using the index $R_\textrm{HK}$ to include the photosphere flux, we find
\begin{equation}
\log{R^\prime_\textrm{FUV}} = (1.87 \pm 0.07) \log{R_\textrm{HK}} + (3.30 \pm 0.33)
\end{equation}
This relation has a similar slope to the slopes \citet{activityvsactivity} found for four prominent transition region lines, even though only part of the FUV flux originates in the transition region. 


Our work in the 1350-1780\AA\ range fills a gap in previous studies of the evolution of stellar activity with age. \citet{sunasastar} presented power-law fits to the decay of X-ray and ultraviolet activity with age among solar analogs. They noted that between the ages of 100~Myr and 6.7~Gyr relatively long-wavelength emission falls off much more slowly with age than shorter-wavelength emission: soft X-ray (0.6-12~keV) emission falls off as $\tau^{-1.92}$, extreme UV (100-360\AA) emission falls off as $\tau^{-1.20}$, and far-UV (920-1180\AA) emission falls off as $\tau^{-0.85}$. Our Equation~\ref{mg_ruvfit} implies that between 30~Myr and 1~Gyr 1350-1780\AA\ emission falls off as $\tau^{-0.42 \pm 0.18}$, continuing the trend although with large scatter. We caution, however, that because we could not estimate $R^\prime_\textrm{FUV}$ for stars of cooler spectral type than K5 this result is based primarily on the AB~Dor moving group and the Hyades cluster, as these groups contributed the bulk of the G- and K-type stars in our sample. 
We also note that Equation~\ref{hip_ruvfit} shows that the excess FUV and excess Ca~II fluxes are proportional to each other, suggesting that the trend of long-wavelength emission decaying more gradually does not extend to longer wavelengths than the FUV.

\subsection{Predictions Among $R^\prime_\textrm{HK}$, Stellar Age, and UV Color Excess}

\subsubsection{The Residuals and their Significance}

The large residuals around our relations between the Ca~II activity indicator $R^\prime_\textrm{HK}$, stellar age, and UV color excess, Equations~\ref{fuvfit}-\ref{nacfit} and \ref{nuvjfit}-\ref{nacjfit}, reflect a genuine limitation of the data. Stars with very different Ca~II fluxes can show similar UV fluxes, and vice versa, as suggested by Figures~\ref{uvacplots_hip}-\ref{acuvplots_hip} and \ref{uvacplots_mg}-\ref{acuvplots_mg}. To show that the scatter is not an artifact of our fitting methods, or even of the decision to fit curves in the first place, we select from the Hipparcos sample pairs of stars whose colors are identical within $1\sigma$ confidence:
\begin{eqnarray}\label{analogdefn}
\left[\frac{\left((FUV-V)_2 - (FUV-V)_1\right)^2}{\sigma_\textrm{FUV,1}^2+\sigma_\textrm{V,1}^2+\sigma_\textrm{FUV,2}^2+\sigma_\textrm{V,2}^2}\right]
 + \left[\frac{\left((B-V)_2 - (B-V)_1\right)^2}{\sigma_\textrm{B-V,1}^2+\sigma_\textrm{B-V,2}^2}\right] < 1 
\qquad \textrm{and} \nonumber \\ 
\left[\frac{\left((NUV-V)_2 - (NUV-V)_1\right)^2}{\sigma_\textrm{NUV,1}^2+\sigma_\textrm{V,1}^2+\sigma_\textrm{NUV,2}^2+\sigma_\textrm{V,2}^2}\right]
 + \left[\frac{\left((B-V)_2 - (B-V)_1\right)^2}{\sigma_\textrm{B-V,1}^2+\sigma_\textrm{B-V,2}^2}\right] < 1
\end{eqnarray}
where $\sigma$ denotes the formal uncertainty on the color or magnitude, and the subscripts 1 and 2 refer to the two stars in each pair. Our Hipparcos sample of 1204 stars allows for 724,206 possible pairings. Of these, 1875 meet the criterion in Equation~\ref{analogdefn}. We will refer to these 1875 pairs of stars as photometric analog pairs. Some stars are members of multiple analog pairs, i.e. their colors resemble those of more than one other star. Analog pairs typically differ in $B-V$ color by a median of 0.017~mag, in $FUV-V$ color by 0.14~mag, and in $NUV-V$ color by 0.016~mag.

\begin{figure*}
\includegraphics[width=0.48\textwidth]{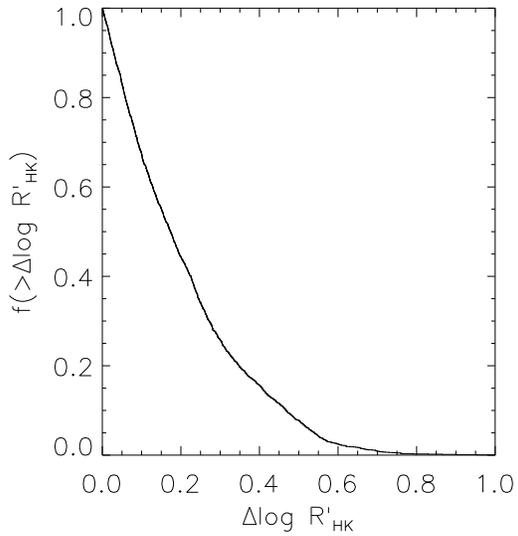}
\caption{The fraction of pairs of photometric analogs differing by more than the amount along the abscissa in $\Delta \log{R^\prime_\textrm{HK}}$. Although 26\% have identical $\log{R^\prime_\textrm{HK}}$ values within the typical scatter of 0.08~dex, 26\% differ by more than 0.3~dex, and 16\% differ by more than 0.4~dex. Because such large differences in $\log{R^\prime_\textrm{HK}}$ are so common among photometrically indistinguishable stars, any attempt to predict $\log{R^\prime_\textrm{HK}}$ from photometry alone will show large residuals, no matter what algorithm is used to make the predictions.}\label{analogrhk}
\end{figure*}

Since the two stars in each analog pair have statistically indistinguishable photometry, one might hope that they also have similar $R^\prime_\textrm{HK}$ measurements. We show the actual distribution of $\left|\log{R^\prime_\textrm{HK,2}}-\log{R^\prime_\textrm{HK,1}}\right|$ for all 1875 pairs in Figure~\ref{analogrhk}. A quarter of the pairs indeed have identical $\log{R^\prime_\textrm{HK}}$ values to within our observed scatter of 0.08~dex, but another quarter of the pairs have highly discrepant $\log{R^\prime_\textrm{HK}}$ values differing by $\sim$ 0.3-0.5~dex. The median difference in $\log{R^\prime_\textrm{HK}}$ is 0.17~dex, close to our RMS residuals of 0.18~dex around Equation~\ref{nacfit}. We have not performed a fit of any kind to arrive at this number; we have simply picked out photometrically identical stars from the data and asked whether their $R^\prime_\textrm{HK}$ measurements are also consistent with each other.

The predictive power of all six of our fits -- Equations~\ref{fuvfit}-\ref{nacfit} and \ref{nuvjfit}-\ref{nacjfit} -- is limited by these large residuals.
Equations~\ref{fuvfit}, \ref{nuvfit}, and \ref{nuvjfit}, which predict the UV fluxes of stars based on their $R^\prime_\textrm{HK}$ values and optical or near-infrared colors, have residuals of 0.33~mag, 0.21~mag, and 0.46~mag, respectively. If we assume that $FUV-V$ for all stars with similar $R^\prime_\textrm{HK}$ follows a Gaussian distribution with a standard deviation equal to our 0.33~mag residuals, we would report a 95\% ($2\sigma$) confidence interval for a star's FUV magnitude that allows nearly a factor of 2 variation in flux. 
These predictions, uncertain as they are, are still much more precise than those we would get were we to ignore stellar activity. If we repeat the fitting procedure for Equations~\ref{fuvfit}, \ref{nuvfit}, and \ref{nuvjfit}, but require that the models have no dependence on $R^\prime_\textrm{HK}$ or age, we find much larger residuals of 0.68~mag, 0.25~mag, and 0.64~mag, respectively.

While Equations~\ref{fuvfit}, \ref{nuvfit}, and \ref{nuvjfit} allow factor of 2 predictions in UV flux, the residuals around Equations~\ref{facfit}, \ref{nacfit}, and \ref{nacjfit} (0.15~dex, 0.18~dex, and 0.39~dex, respectively) prevent us from using UV fluxes to predict $R^\prime_\textrm{HK}$ to better precision than half its observed range, or stellar age to better than an order of magnitude. 
Consider as an example a star with $B-V = 0.65$ and $FUV-V = 11.25$, near the UV-bright edge of the stellar locus in Figure~\ref{ccplots_hip}. Equation~\ref{facfit} states that such stars have $\log{R^\prime_\textrm{HK}} = -4.48$ on average. If we assume that $\log{R^\prime_\textrm{HK}}$ for all stars with similar colors follows a Gaussian distribution with a standard deviation equal to our RMS residuals of 0.15~dex, we report a 95\% confidence interval of $-4.78 < \log{R^\prime_\textrm{HK}} < -4.19$. This is an enormous range in activity, corresponding to an age anywhere between 50~Myr and 3~Gyr \citep{chro_levels}. Even though the star has a very high UV flux compared to other stars of the same $B-V$, we can infer little more than that the star is more active than the Sun. 
We found a similar result in \citet{tausco}, where we selected active stars by their UV flux but found that only a third showed obvious Ca~II or H$\alpha$ emission in follow-up optical spectra. The remainder, like the 3~Gyr stars in the preceding example, were not unusually active.

\subsubsection{Systematics in Our Fits}

Equations~\ref{fuvfit}-\ref{nacfit} can give counterintuitive results if used carelessly, because they are best-fit models in the nonparametric sense. Instead of specifying a hypothetical one-to-one relationship between UV flux and $\log{R^\prime_\textrm{HK}}$, they give the average $UV-V$ in a large sample of stars with the same measured values of $B-V$ and $R^\prime_\textrm{HK}$, or the average $\log{R^\prime_\textrm{HK}}$ in a large sample of stars with the same measured values of $B-V$ and $UV-V$. The average is, implicitly, over a population with the same $R^\prime_\textrm{HK}$ distribution as our sample. The log-age predictions of Equations~\ref{nuvjfit}-\ref{nacjfit} must be treated as averages in the same sense.
This averaging process is unusually significant for our fits because they have large residuals, and most important for Equations~\ref{facfit}, \ref{nacfit}, and \ref{nacjfit}, where the residuals are a significant fraction of the width of the parameter space. 

One consequence of the fits acting as averages is visible in Figure~\ref{acuvplots_mg}. Equation~\ref{nacjfit} systematically under-predicts the age of the oldest stars in the moving group and cluster sample, those in the Hyades, and systematically over-predicts the age of the youngest stars, those in the TW~Hydrae association. 
The reason can be seen with a thought experiment. A particular (high) observed UV flux can be produced by a very young star, such as a TW~Hydrae member, with UV emission typical for its age, or it can be produced by a moderately young star, such as a Tucana/Horologium or AB~Doradus member, that is (or is measured to be) UV-bright for its age. It \emph{cannot} be produced by an even younger star that appears UV-faint for its age, because TW~Hydrae is the youngest association in our sample.
The mean age predicted by Equation~\ref{nacjfit} for a star with the UV flux typical of a TW~Hya member will be older than the age of TW~Hya, because there are older stars with the same flux but no younger stars. More generally, the ages predicted by Equation~\ref{nacjfit} are biased away from the edges of the $10^{6.9}-10^{8.8}$~Myr age range probed by our moving group and cluster sample. 

The edge effect we identified for ages is much weaker for the $R^\prime_\textrm{HK}$ values predicted by Equations~\ref{facfit} and \ref{nacfit}, because the Hipparcos sample is volume limited so the activity distribution in the sample approximates that in the Galactic stellar population. 
However, the most extremely active ($\log{R^\prime_\textrm{HK}} \gtrsim -4.0$) or inactive ($\log{R^\prime_\textrm{HK}} \lesssim -5.5$) stars are not represented, so we expect Equations~\ref{facfit} and \ref{nacfit} will slightly underpredict the $R^\prime_\textrm{HK}$ of the most active stars and overpredict that of the least active stars. 

\label{nopreselect}

Other systematic effects may also skew the predictions at very high or very low activity levels.
We 
tested the predictive power of Equations~\ref{facfit} and \ref{nacfit} on the activity sample of \citet{sindex_duncan}. They gave Ca~II fluxes, in the form of the Mount Wilson $S$-index, for 12 stars that were detected by GALEX but were too faint to include in the Hipparcos catalog. Because they are not Hipparcos stars, these 12 stars form a sample independent of our calibration data. 8 of these stars meet the requirements of either Equation~\ref{facfit} or \ref{nacfit}. We found 95\% confidence intervals for $\log{R^\prime_\textrm{HK}}$ based on the GALEX data, approximating the scatter around our fits as Gaussian in $\log{R^\prime_\textrm{HK}}$. In Table~\ref{duncantest}, we compare our confidence intervals with $\log{R^\prime_\textrm{HK}}$ values we computed from the published $S$-index measurements following the prescription of \citet{rhkcalc}.
Only 5 of the 8 stars' $\log{R^\prime_\textrm{HK}}$ measurements lie within our 95\% confidence interval, while we expect at least 7 out of 8 to do so. Since the scatter around the fits is close to Gaussian, as assumed, it follows that there are as yet unidentified systematics in the fits themselves. We note that the three discrepant stars include the two most active stars in the sample, with $\log{R^\prime_\textrm{HK}} > -4.0$, suggesting that very active stars may be the problem area.

\subsection{Unaccounted-For Variables}

In this study, we have ignored metallicity. We expect metallicity to affect stellar UV luminosities in three ways. 
First, we expect the photospheres of metal-rich stars to have higher UV opacities, particularly from line blanketing, and redder UV-optical colors compared to metal-poor stars \citep{gray_atmospheres}. 
Second, FUV emission comes from metal lines and continuum, so chromospheric and transition region gas around more metal-rich stars should produce more FUV flux per unit emission measure. 
Finally, metal line emission is the primary cooling mechanism for transition region gas, so we expect the structure of the outer stellar atmosphere -- in particular, the thickness and density of the transition region -- to depend on the stellar metallicity, with potentially complex effects on the observed activity. The number of processes to untangle, and our inability to thoroughly sample a large metallicity range, preclude our study of the metallicity-dependence of stellar UV flux.

We have also ignored rotation, even though activity-rotation relations are, on theoretical grounds, more fundamental than activity-age relations. Stellar activity results from chromospheric and coronal heating, which among Sun-like stars is mediated by the stellar magnetic field. Every major model for the solar dynamo incorporates differential rotation in some form \citep{solarmodel_review}, and so the strength of the magnetic field might be expected to scale with the overall rotation rate of the star. Activity should therefore be strongly correlated with rotation, whereas the relationship between activity and age represents the combination of the activity-rotation correlation with the systematic slowing of stellar rotation with age. \citet{chro_levels} confirm that activity-rotation-age relations show a tighter correlation than activity-age relations that do not incorporate measured rotation periods. However, rotation periods are not available for as many stars as $R^\prime_\textrm{HK}$, and are not even detectable in all stars surveyed for rotation \citep[e.g.][]{rot_example}.

While we expect stellar UV flux to be affected by both metallicity and rotation, we do not need to invoke either factor to explain our data. The scatter we observe in UV flux is consistent with that expected from propagated errors and previously measured variability, so without more precise photometry and coeval activity measurements the data do not support adding more variables to the analysis. We have instead restricted the problem to its three essential dimensions: UV flux, photosphere effective temperature, and one age or activity indicator to which we may compare the UV flux.

\section{Summary}

We set out to quantify the dependence of stellar ultraviolet broad-band flux on age among Sun-like stars, and to test the correlation of UV flux with more traditional activity indicators, in particular the well studied chromospheric indicator $R^\prime_\textrm{HK}$. To this end, we matched both an unbiased, volume-limited sample of nearby stars with $R^\prime_\textrm{HK}$ measurements but unknown ages, and an age-calibrated but incomplete sample of young stars, to archived GALEX data. To avoid introducing a model dependence into our results, we used nonparametric fits to describe the correlation between UV-optical or UV-infrared colors and $R^\prime_\textrm{HK}$ or age measurements. We also explored the construction of model-dependent activity indicators, $R^\prime_\textrm{FUV}$ and $R^\prime_\textrm{NUV}$, defined analogously to $R^\prime_\textrm{HK}$.

Our major results may be summarized as follows:
\begin{itemize}
\item We detect a clear correlation between UV color and both $R^\prime_\textrm{HK}$ and age, in the sense that younger stars with higher $R^\prime_\textrm{HK}$ have higher UV luminosities. We provide relations describing the average $R^\prime_\textrm{HK}$ of stars at fixed UV color, and the average UV color of stars at fixed $R^\prime_\textrm{HK}$, where both averages are understood to be taken over a volume-limited sample. We likewise construct relations describing the average age of stars at fixed UV color, and the average UV color of stars at fixed age, although given the heterogeneity of the sample on which these latter relations are based the meaning of the averages is not as clear.
\item Our mean relations are not good predictors of \emph{individual} $R^\prime_\textrm{HK}$ values ($\pm 0.15-0.18$~dex) or ages ($\pm 0.4$~dex, with additional systematics). Selecting young, active stars using UV excess techniques requires a thorough spectroscopic follow-up campaign that is likely to reject many UV-selected targets as false positives \citep[cf.][]{tausco}. Variability and measurement error appear to be equally responsible for the scatter around our mean relations.
\item We find that attempts to define $R^\prime_\textrm{FUV}$ or $R^\prime_\textrm{NUV}$ indices are dominated by systematic uncertainties in the photosphere contribution, particularly in the NUV. Until these uncertainties can be constrained observationally, use of such indices is likely premature. We recommend instead a purely empirical activity index such as UV-optical color, despite the strong temperature-dependence such indices have.
\item We find that the FUV excess flux between 1350\AA\ and 1780\AA\ is proportional to the excess flux in the Ca~II H and K lines. We tentatively find that the FUV excess flux decays with time as $\tau^{-0.42 \pm 0.18}$, extending the results of \citet{sunasastar} to longer wavelengths. This decay rate should be confirmed with data that sample a wider variety of ages.
\end{itemize}

Our results on the age dependence of the FUV flux are currently limited by the lack of young stars of known age with GALEX data. Improving these results will likely require expanded moving group membership lists -- most open clusters near enough to probe down to K-type photospheres at AIS survey depth have at least one star bright enough to pose a danger to GALEX, forcing GALEX to either observe only the outskirts of the cluster or avoid it entirely. Acquiring more precise near-infrared photometry than the often saturated 2MASS fluxes we used would also improve many of our age results, even if the stellar sample remained unchanged.

The GALEX All Sky Survey contains 111,755,312 sources\footnote{Based on an SQL query performed at \url{http://galex.stsci.edu/GR6/}}, of which $\sim 80\%$ are stars \citep{2005ApJ...619L..27B}. 
Far more stars already have ultraviolet photometry than will have spectroscopic measurements in the foreseeable future. We have shown that ultraviolet photometry does not show a tight correlation to optical activity indicators such as $R^\prime_\textrm{HK}$, and cannot be used to cleanly identify high-$R^\prime_\textrm{HK}$ stars. However, it can still be used to filter a sample, removing the most inactive stars to allow more efficient follow-up of the rest. No other activity indicator is as readily available as FUV and NUV photometry.

\acknowledgments

We would like to thank 
the referee for many insightful comments and suggestions, 
Ted Wyder and Patrick Morrissey for addressing our questions about the GALEX data, 
and 
Chad Schafer and John Carpenter for their advice on the statistical analysis. 
This research was supported in part by NASA grant NNX08AH95G to L.A.H.



{\it Facilities:} \facility{GALEX}, \facility{HIPPARCOS}, \facility{KPNO:CFT}

\appendix

\section{Optimal Photosphere Colors for UV Studies}\label{bestcolors}

As we noted in \citet{tausco}, the precision to which UV photometry can be used as an activity measure is limited by the precision to which we know the photosphere UV flux. The uncertainty in the photosphere flux, in turn, is dominated by the precision to which the star can be classified because of the steep dependence of UV flux on spectral type. We consider here the use of a single color for stellar classification, implicit in the use of color-color diagrams such as Figure~\ref{ccplots_hip}. This analysis does not apply to the multi-band fits we used to derive $R^\prime_\textrm{UV}$ in Section~\ref{rprimeuv}. However, it would apply to an attempt to use the $B-V$ color to estimate both $m_\textrm{bol}$ and the photospheric UV flux, and derive $R^\prime_\textrm{UV}$ from these single-color values.

The uncertainty in the UV flux is roughly
\begin{eqnarray}
\sigma_\textrm{UV} & \approx & \left(\frac{\partial (UV-X)}{\partial C}\right) \sigma_C \nonumber \\
 & \sim & \left(\frac{\Delta (UV-X)}{\Delta C}\right) \sigma_C \nonumber \\
 & = & \Delta (UV-X) \frac{\sigma_C}{\Delta C} \label{uvsig}
\end{eqnarray}
where the color $UV-X$ represents the UV flux relative to flux in a purely photospheric band, $C$ is the color used to classify the star, and $\Delta$ denotes the range of a quantity observed along the main sequence.
Minimizing the first factor in Equation~\ref{uvsig} involves choosing a diagnostic color $UV-X$ that varies relatively little with spectral type, while minimizing the second involves choosing a photosphere color $C$ that resolves the main sequence into as many elements as possible. 

$\Delta(UV-X)$ is smallest if $X$ is as blue a band as possible. In particular, an analysis based on $UV-B$ or $UV-V$ color will always be more precise than one based on $UV-J$ or $UV-K$. We adopted $UV-V$ rather than $UV-B$ because we could determine stars' $V$ magnitudes much more precisely, and because the penalty Equation~\ref{uvsig} predicts for using $V$, $\frac{\Delta (FUV-V)}{\Delta (FUV-B)} \sim \frac{\Delta (NUV-V)}{\Delta (NUV-B)} \sim 1.1$, is small.

Minimizing the classification term $\sigma_C/\Delta C$ is more complex, because it depends on the quality of the available photometry and not just on the position of the stellar locus in color space. We found $\sigma_C$ and $\Delta C$ for several colors in our Hipparcos sample, adopting the median formal error for $\sigma$ and adopting the range between the 10th and 90th percentiles for $\Delta$ to avoid biases from outliers. We found $\sigma_{B-V}/\Delta (B-V) = 0.04$, $\sigma_{J-K}/\Delta (J-K) = 0.35$, and $\sigma_{V-K}/\Delta (V-K) = 0.08$. Infrared colors are a poor classifier for our sample because many of our target stars are saturated in 2MASS, giving large ($\sim 0.2$~mag) errors in their near infrared magnitudes. As a result, we identified stars by their $B-V$ color for the Hipparcos sample, even though using $J-K$ would allow us to compare directly to our results for the moving group and cluster sample, or to the measurements we obtained in \citet{tausco}.

\section{Expected Contributions to the FUV Flux}\label{fuvfrac}

\citet{alphacen} present a detailed STIS spectrum of $\alpha$~Cen~A, including a complete line list (their Table~4) and continuum estimates (dashed line in their Figure~2). We use this spectrum as a template to estimate the contribution to the FUV flux from various sources in Sun-like stars.

Adding together the fluxes of all the lines between 1350\AA\ and 1690\AA, we find $\alpha$~Cen~A has a total FUV line flux\footnote{We assume the flux units in Table~4 are $10^{-15}$~erg~s$^{-1}$~cm$^{-2}$ rather than the claimed mW~m$^{-2}$, as the former units make the tabulated fluxes consistent with both the surface fluxes listed in \citet{alphacen} Table~7 and with the flux densities in their Figure~2.} of $2.3 \times 10^{-11}$~erg~s$^{-1}$~cm$^{-2}$. The continuum seen in \citet{alphacen} Figure~2 varies between $5 \times 10^{-12}$~erg~s$^{-1}$~cm$^{-2}$~\AA$^{-1}$ at 1370\AA\ and $3 \times 10^{-13}$~erg~s$^{-1}$~cm$^{-2}$~\AA$^{-1}$ at 1690\AA. Assuming an average continuum of $1.5 \times 10^{-13}$~erg~s$^{-1}$~cm$^{-2}$~\AA$^{-1}$, we infer a total continuum flux of $5.1 \times 10^{-11}$~erg~s$^{-1}$~cm$^{-2}$ over the range 1350-1690\AA. It follows that, for a typical Sun-like star, lines contribute 30\% of the GALEX FUV flux, with the rest coming from a weak continuum. It is possible that even at the high resolution of the STIS observations ($R \sim 114,000$) some of the ``continuum'' consists of unresolved lines, but this has not been investigated (Pagano, priv. comm.).

Although the Lyman $\alpha$ line in their spectrum is heavily contaminated with both interstellar absorption and geocoronal emission, \citet{alphacen} estimated a flux of $1.04 \times 10^{-10}$~erg~s$^{-1}$~cm$^{-2}$ by fitting the wings of the line profile. In comparison, the total line flux over the range covered by the observations, 1170-1690\AA, is $1.39 \times 10^{-10}$~erg~s$^{-1}$~cm$^{-2}$, and our estimate for the continuum in that range is $7.8 \times 10^{-11}$~erg~s$^{-1}$~cm$^{-2}$. Therefore, Lyman $\alpha$ contributes 75\% of the line flux and 48\% of the total flux in the range 1170-1690\AA.

\bibliographystyle{hapj}
\bibliography{phantoms,../../references}

\begin{thebibliography}{39}
\expandafter\ifx\csname natexlab\endcsname\relax\def\natexlab#1{#1}\fi

\bibitem[{Aitkin(1981)}]{fitylimits}
Aitkin, M. 1981, Technometrics, 23, 161

\bibitem[{{Bessell}(1990{\natexlab{a}})}]{landolt_vega}
{Bessell}, M.~S. 1990{\natexlab{a}}, \aaps, 83, 357

\bibitem[{{Bessell}(1990{\natexlab{b}})}]{bessellcurves}
------. 1990{\natexlab{b}}, \pasp, 102, 1181

\bibitem[{{Bianchi} {et~al.}(2005){Bianchi}, {Seibert}, {Zheng}, {Thilker},
  {Friedman}, {Wyder}, {Donas}, {Barlow}, {Byun}, {Forster}, {Heckman},
  {Jelinsky}, {Lee}, {Madore}, {Malina}, {Martin}, {Milliard}, {Morrissey},
  {Neff}, {Rich}, {Schiminovich}, {Siegmund}, {Small}, {Szalay}, \&
  {Welsh}}]{2005ApJ...619L..27B}
{Bianchi}, L. {et~al.} 2005, \apjl, 619, L27

\bibitem[{{Browne} {et~al.}(2009){Browne}, {Welsh}, \&
  {Wheatley}}]{galexpleiades}
{Browne}, S.~E., {Welsh}, B.~Y., \& {Wheatley}, J. 2009, \pasp, 121, 450,
  astro-ph/0904.0042

\bibitem[{{Cardelli} {et~al.}(1989){Cardelli}, {Clayton}, \&
  {Mathis}}]{av_curve}
{Cardelli}, J.~A., {Clayton}, G.~C., \& {Mathis}, J.~S. 1989, \apj, 345, 245

\bibitem[{{Cargile} {et~al.}(2009){Cargile}, {James}, \&
  {Platais}}]{blancoreddening}
{Cargile}, P.~A., {James}, D.~J., \& {Platais}, I. 2009, \aj, 137, 3230,
  astro-ph/0901.2368

\bibitem[{{Charbonneau}(2010)}]{solarmodel_review}
{Charbonneau}, P. 2010, Living Reviews in Solar Physics, 7, 3

\bibitem[{{Charbonneau} \& {MacGregor}(1993)}]{wind_braking}
{Charbonneau}, P., \& {MacGregor}, K.~B. 1993, \apj, 417, 762

\bibitem[{{Donahue} {et~al.}(1996){Donahue}, {Saar}, \&
  {Baliunas}}]{rot_example}
{Donahue}, R.~A., {Saar}, S.~H., \& {Baliunas}, S.~L. 1996, \apj, 466, 384

\bibitem[{{Duncan} {et~al.}(1991){Duncan}, {Vaughan}, {Wilson}, {Preston},
  {Frazer}, {Lanning}, {Misch}, {Mueller}, {Soyumer}, {Woodard}, {Baliunas},
  {Noyes}, {Hartmann}, {Porter}, {Zwaan}, {Middelkoop}, {Rutten}, \&
  {Mihalas}}]{sindex_duncan}
{Duncan}, D.~K. {et~al.} 1991, \apjs, 76, 383

\bibitem[{{Findeisen} \& {Hillenbrand}(2010)}]{tausco}
{Findeisen}, K., \& {Hillenbrand}, L. 2010, \aj, 139, 1338, astro-ph/1001.3684

\bibitem[{{Gorti} \& {Hollenbach}(2009)}]{diskevol1}
{Gorti}, U., \& {Hollenbach}, D. 2009, \apj, 690, 1539, astro-ph/0809.1494

\bibitem[{{Gray}(2005)}]{gray_atmospheres}
{Gray}, D.~F. 2005, {The Observation and Analysis of Stellar Photospheres}, 3rd
  edn. (Cambridge University Press)

\bibitem[{{Guillout} {et~al.}(1999){Guillout}, {Schmitt}, {Egret}, {Voges},
  {Motch}, \& {Sterzik}}]{rosatefficiency}
{Guillout}, P., {Schmitt}, J.~H.~M.~M., {Egret}, D., {Voges}, W., {Motch}, C.,
  \& {Sterzik}, M.~F. 1999, \aap, 351, 1003

\bibitem[{{Henry} {et~al.}(1996){Henry}, {Soderblom}, {Donahue}, \&
  {Baliunas}}]{1996AJ....111..439H}
{Henry}, T.~J., {Soderblom}, D.~R., {Donahue}, R.~A., \& {Baliunas}, S.~L.
  1996, \aj, 111, 439

\bibitem[{{Isaacson} \& {Fischer}(2010)}]{sindex_isaacson}
{Isaacson}, H., \& {Fischer}, D. 2010, \apj, 725, 875, astro-ph/1009.2301

\bibitem[{{Jester} {et~al.}(2005){Jester}, {Schneider}, {Richards}, {Green},
  {Schmidt}, {Hall}, {Strauss}, {Vanden Berk}, {Stoughton}, {Gunn},
  {Brinkmann}, {Kent}, {Smith}, {Tucker}, \& {Yanny}}]{kh_bv}
{Jester}, S. {et~al.} 2005, \aj, 130, 873, arXiv:astro-ph/0506022

\bibitem[{{Kraus} \& {Hillenbrand}(2007)}]{mscolors}
{Kraus}, A.~L., \& {Hillenbrand}, L.~A. 2007, \aj, 134, 2340, 0708.2719

\bibitem[{{Mamajek} \& {Hillenbrand}(2008)}]{chro_levels}
{Mamajek}, E.~E., \& {Hillenbrand}, L.~A. 2008, \apj, 687, 1264,
  astro-ph/0807.1686

\bibitem[{{Martin} {et~al.}(2005){Martin}, {Fanson}, {Schiminovich},
  {Morrissey}, {Friedman}, {Barlow}, {Conrow}, {Grange}, {Jelinsky},
  {Milliard}, {Siegmund}, {Bianchi}, {Byun}, {Donas}, {Forster}, {Heckman},
  {Lee}, {Madore}, {Malina}, {Neff}, {Rich}, {Small}, {Surber}, {Szalay},
  {Welsh}, \& {Wyder}}]{galexintro}
{Martin}, D.~C. {et~al.} 2005, \apjl, 619, L1, arXiv:astro-ph/0411302

\bibitem[{{Mermilliod} {et~al.}(2008){Mermilliod}, {Platais}, {James},
  {Grenon}, \& {Cargile}}]{blancomembers}
{Mermilliod}, J., {Platais}, I., {James}, D.~J., {Grenon}, M., \& {Cargile},
  P.~A. 2008, \aap, 485, 95

\bibitem[{{Morrissey} {et~al.}(2007){Morrissey}, {Conrow}, {Barlow}, {Small},
  {Seibert}, {Wyder}, {Budav{\'a}ri}, {Arnouts}, {Friedman}, {Forster},
  {Martin}, {Neff}, {Schiminovich}, {Bianchi}, {Donas}, {Heckman}, {Lee},
  {Madore}, {Milliard}, {Rich}, {Szalay}, {Welsh}, \& {Yi}}]{galexcalib}
{Morrissey}, P. {et~al.} 2007, ApJ, 173, S682

\bibitem[{{Noyes} {et~al.}(1984){Noyes}, {Hartmann}, {Baliunas}, {Duncan}, \&
  {Vaughan}}]{rhkcalc}
{Noyes}, R.~W., {Hartmann}, L.~W., {Baliunas}, S.~L., {Duncan}, D.~K., \&
  {Vaughan}, A.~H. 1984, \apj, 279, 763

\bibitem[{{Oke}(1974)}]{abmags}
{Oke}, J.~B. 1974, \apjs, 27, 21

\bibitem[{{Pagano} {et~al.}(2004){Pagano}, {Linsky}, {Valenti}, \&
  {Duncan}}]{alphacen}
{Pagano}, I., {Linsky}, J.~L., {Valenti}, J., \& {Duncan}, D.~K. 2004, \aap,
  415, 331, astro-ph/0310901

\bibitem[{{Perryman} {et~al.}(1998){Perryman}, {Brown}, {Lebreton}, {Gomez},
  {Turon}, {Cayrel de Strobel}, {Mermilliod}, {Robichon}, {Kovalevsky}, \&
  {Crifo}}]{hyamembers}
{Perryman}, M.~A.~C. {et~al.} 1998, \aap, 331, 81, astro-ph/9707253

\bibitem[{{Preibisch} \& {Feigelson}(2005)}]{redyouth}
{Preibisch}, T., \& {Feigelson}, E.~D. 2005, \apjs, 160, 390, astro-ph/0506052

\bibitem[{{Ribas} {et~al.}(2005){Ribas}, {Guinan}, {G{\"u}del}, \&
  {Audard}}]{sunasastar}
{Ribas}, I., {Guinan}, E.~F., {G{\"u}del}, M., \& {Audard}, M. 2005, \apj, 622,
  680, astro-ph/0412253

\bibitem[{{Rutten} {et~al.}(1991){Rutten}, {Schrijver}, {Lemmens}, \&
  {Zwaan}}]{activityvsactivity}
{Rutten}, R.~G.~M., {Schrijver}, C.~J., {Lemmens}, A.~F.~P., \& {Zwaan}, C.
  1991, \aap, 252, 203

\bibitem[{{Shkolnik} {et~al.}(2011){Shkolnik}, {Liu}, {Reid}, {Dupuy}, \&
  {Weinberger}}]{galexmdwarfs}
{Shkolnik}, E.~L., {Liu}, M.~C., {Reid}, I.~N., {Dupuy}, T., \& {Weinberger},
  A.~J. 2011, \apj, 727, 6, astro-ph/1011.2708

\bibitem[{{Simon} {et~al.}(1985){Simon}, {Herbig}, \&
  {Boesgaard}}]{1985ApJ...293..551S}
{Simon}, T., {Herbig}, G., \& {Boesgaard}, A.~M. 1985, \apj, 293, 551

\bibitem[{{Soderblom}(1985)}]{sindex_error}
{Soderblom}, D.~R. 1985, \aj, 90, 2103

\bibitem[{{Soderblom} {et~al.}(1991){Soderblom}, {Duncan}, \&
  {Johnson}}]{soderblom91}
{Soderblom}, D.~R., {Duncan}, D.~K., \& {Johnson}, D.~R.~H. 1991, \apj, 375,
  722

\bibitem[{{Valenti} {et~al.}(2000){Valenti}, {Johns-Krull}, \&
  {Linsky}}]{valenti}
{Valenti}, J.~A., {Johns-Krull}, C.~M., \& {Linsky}, J.~L. 2000, \apjs, 129,
  399

\bibitem[{{Vaughan} \& {Preston}(1980)}]{vpsurvey}
{Vaughan}, A.~H., \& {Preston}, G.~W. 1980, \pasp, 92, 385

\bibitem[{Wasserman(2006)}]{crossvalidation}
Wasserman, L. 2006, All of Nonparametric Statistics (New York: Springer
  Science+Business Media)

\bibitem[{{Wright} {et~al.}(2004){Wright}, {Marcy}, {Butler}, \&
  {Vogt}}]{sindex_wright}
{Wright}, J.~T., {Marcy}, G.~W., {Butler}, R.~P., \& {Vogt}, S.~S. 2004, \apjs,
  152, 261, astro-ph/0402582

\bibitem[{{Zuckerman} \& {Song}(2004)}]{mgreview_old}
{Zuckerman}, B., \& {Song}, I. 2004, \araa, 42, 685

\end{thebibliography}

\clearpage

\begin{deluxetable}{c c c c c c c c c}
\tablewidth{0pt}
\tablecaption{Magnitudes Used to Fit Photosphere Fluxes to Observed Stars\label{modelmags}}
\tablehead{\colhead{Spectral Type} & \colhead{$M_\textrm{FUV}$} & \colhead{$M_\textrm{NUV}$} & \colhead{$M_\textrm{B}$} & \colhead{$M_\textrm{V}$} & \colhead{$M_\textrm{J}$} & \colhead{$M_\textrm{H}$} & \colhead{$M_\textrm{K}$} & \colhead{$M_\textrm{bol}$}}
\startdata
B8 &  $0.80$ &  $0.57$ & $-0.31$ & $-0.20$ & $0.01$ & $0.10$ & $0.11$ & $-1.00$	\\
A0 &  $3.08$ &  $2.26$ &  $0.59$ &  $0.61$ & $0.54$ & $0.58$ & $0.56$ &  $0.30$	\\
A2 &  $5.05$ &  $3.30$ &  $1.36$ &  $1.31$ & $1.12$ & $1.15$ & $1.12$ &  $1.10$	\\
A5 &  $7.42$ &  $4.31$ &  $2.06$ &  $1.91$ & $1.53$ & $1.52$ & $1.48$ &  $1.75$	\\
A7 &  $8.44$ &  $4.83$ &  $2.41$ &  $2.21$ & $1.75$ & $1.71$ & $1.66$ &  $2.08$	\\
F0 & $10.50$ &  $5.83$ &  $3.02$ &  $2.71$ & $2.10$ & $2.01$ & $1.96$ &  $2.61$	\\
F2 & $11.27$ &  $6.29$ &  $3.35$ &  $3.01$ & $2.32$ & $2.20$ & $2.14$ &  $2.89$	\\
F5 & $13.41$ &  $7.46$ &  $4.17$ &  $3.76$ & $2.85$ & $2.67$ & $2.61$ &  $3.61$	\\
F8 & $16.35$ &  $8.77$ &  $4.93$ &  $4.41$ & $3.31$ & $3.08$ & $3.01$ &  $4.24$	\\
G0 & $17.65$ &  $9.40$ &  $5.24$ &  $4.66$ & $3.53$ & $3.27$ & $3.20$ &  $4.47$	\\
G2 & $18.65$ &  $9.88$ &  $5.44$ &  $4.81$ & $3.64$ & $3.38$ & $3.30$ &  $4.60$	\\
G5 & $19.74$ & $10.57$ &  $5.80$ &  $5.11$ & $3.86$ & $3.56$ & $3.48$ &  $4.89$	\\
G8 & $21.11$ & $11.56$ &  $6.45$ &  $5.71$ & $4.31$ & $3.95$ & $3.86$ &  $5.30$	\\
K0 & $22.66$ & $12.53$ &  $6.83$ &  $6.01$ & $4.49$ & $4.10$ & $4.00$ &  $5.69$	\\
K2 & $24.73$ & $14.02$ &  $7.43$ &  $6.51$ & $4.80$ & $4.35$ & $4.24$ &  $6.08$	\\
K4 & $28.23$ & $16.26$ &  $8.16$ &  $7.11$ & $5.08$ & $4.56$ & $4.43$ &  $6.55$	\\
K5 & $30.35$ & $17.72$ &  $8.56$ &  $7.41$ & $5.20$ & $4.64$ & $4.51$ &  $6.68$	\\
\enddata
\tablecomments{FUV and NUV magnitudes are calculated on the GALEX system as described in the text. J, H, K, and bolometric magnitudes are taken from \citet{mscolors} Table~5. B and V magnitudes are derived from the g and r magnitudes in \citet{mscolors} Table~5 using the transformations of \citet{kh_bv}. FUV and NUV magnitudes are not available for K7 or cooler spectral types because Kurucz models do not cover stars with redder $B-V$.}
\end{deluxetable}

\begin{deluxetable}{c c c c c}
\tablewidth{0pt}
\tablecaption{RMS Change in Our Fits Under Alternate Functional Forms\label{deltaf}}
\tablehead{\colhead{Adopted Fit} & \colhead{Residuals} & \colhead{$(\Delta \hat{f})_\textrm{rms}$} & \colhead{$(\Delta \hat{f})_\textrm{rms}$} & \colhead{$(\Delta \hat{f})_\textrm{rms}$} \\
\colhead{(Model 1)} & \colhead{(Model 1)} & \colhead{Model 2 - Model 1} & \colhead{Model 3 - Model 1} & \colhead{Model 4 - Model 1}}
\startdata
Equation~\ref{fuvfit}  & 0.33~mag & 0.033~mag & 0.018~mag & 0.039~mag \\
Equation~\ref{facfit}  & 0.15~dex & 0.035~dex & 0.021~dex & 0.017~dex \\
Equation~\ref{nuvfit}  & 0.21~mag & 0.063~mag & 0.063~mag & 0.051~mag \\
Equation~\ref{nacfit}  & 0.18~dex & 0.038~dex & 0.056~dex & 0.066~dex \\
Equation~\ref{nuvjfit} & 0.46~mag & 0.101~mag & 0.114~mag & 0.122~mag \\
Equation~\ref{nacjfit} & 0.39~dex & 0.092~dex & 0.063~dex & 0.105~dex \\
\enddata
\tablecomments{The RMS difference $\Delta \hat{f}$ between our adopted form for Equations~\ref{fuvfit}-\ref{nacfit} and \ref{nuvjfit}-\ref{nacjfit} and three alternate forms with comparable goodness-of-fit, averaged over the parameter space where each equation is valid. 
Were we to fit a different-degree polynomial for one of these equations, our predictions would be changed by roughly $\Delta \hat{f}$.
Since $\Delta \hat{f}$ is typically much smaller than the residuals around our fits, the final choice of form for Equations~\ref{fuvfit}-\ref{nacfit} and \ref{nuvjfit}-\ref{nacjfit} does not affect our results.}
\end{deluxetable}

\begin{deluxetable}{r c c c}
\tablewidth{0pt}
\tablecaption{Mean Magnitude and Parallax in the Hipparcos Sample\label{biastable}}
\tablehead{\colhead{B-V} & \colhead{$\log{R^\prime_\textrm{HK}} < -5.0$} & \colhead{$-5.0 < \log{R^\prime_\textrm{HK}} < -4.5$} & \colhead{$-4.5 < \log{R^\prime_\textrm{HK}}$}}
\startdata
\cutinhead{Mean V Magnitude}
0.5 -- 0.6 & $6.45 \pm 0.14$ & $7.01 \pm 0.04$ & $7.03 \pm 0.09$ \\
0.6 -- 0.7 & $7.29 \pm 0.08$ & $7.60 \pm 0.04$ & $7.67 \pm 0.08$ \\
0.7 -- 0.8 & $8.06 \pm 0.09$ & $8.07 \pm 0.05$ & $8.32 \pm 0.09$ \\
0.8 -- 0.9 & $8.62 \pm 0.11$ & $8.53 \pm 0.06$ & $8.42 \pm 0.12$ \\
0.9 -- 1.0 & $9.03 \pm 0.16$ & $9.03 \pm 0.06$ & $8.98 \pm 0.10$ \\
\cutinhead{Mean Parallax (mas)}
0.5 -- 0.6 & $32.1 \pm 2.3$ & $29.1 \pm 0.7$ & $29.1 \pm 1.2$ \\
0.6 -- 0.7 & $28.2 \pm 1.0$ & $29.6 \pm 0.6$ & $30.2 \pm 1.2$ \\
0.7 -- 0.8 & $30.4 \pm 1.4$ & $31.8 \pm 1.1$ & $29.5 \pm 1.5$ \\
0.8 -- 0.9 & $32.5 \pm 1.6$ & $32.8 \pm 1.3$ & $36.5 \pm 3.4$ \\
0.9 -- 1.0 & $30.2 \pm 2.0$ & $32.2 \pm 0.9$ & $32.1 \pm 1.3$ \\
\enddata
\tablecomments{Mean $V$ magnitudes and parallaxes, with $1\sigma$ errors on the means, as a function of activity. All pairs of parallaxes are indistinguishable at 95\% confidence. For $B-V < 0.7$, the least active bin is significantly brighter. However, since we see the strongest UV-activity correlations among the reddest stars, the correlation of $V$ with activity among the blue stars cannot be responsible for our trends between UV flux and activity.}
\end{deluxetable}



\begin{deluxetable}{r cc cc c c cc c}
\tablewidth{0pt}
\tabletypesize{\footnotesize}
\tablecaption{Predicted Activity Levels for Selected non-Hipparcos Stars\label{duncantest}}
\tablehead{\colhead{Name} & \colhead{FUV-V} & \colhead{$\sigma_{FUV}$} & \colhead{NUV-V} & \colhead{$\sigma_{NUV}$} & \colhead{B-V} & \colhead{95\% CI} & \colhead{Observed} & \colhead{$\sigma_{\log{R^\prime_\textrm{HK}}}$}\\
\colhead{} & \colhead{(mag)} & \colhead{(mag)} & \colhead{(mag)} & \colhead{(mag)} & \colhead{(mag)} & \colhead{$\log{R^\prime_\textrm{HK}}$} & \colhead{$\log{R^\prime_\textrm{HK}}$} & \colhead{}}
\startdata
BD+01\degr\ 0306 & \nodata & \nodata & 7.65 & 0.01 & 0.96 & $(-5.10, -4.52)$ & $-5.07$ & 0.005 \\
Cl* NGC~2632 KW~127 & 11.17 & 0.13 & \nodata & \nodata & 0.60 & $(-5.01, -4.31)$ & $-4.59$ & \nodata \\
Cl* NGC~2632 KW~217 & 9.93 & 0.04 & \nodata & \nodata & 0.51 & $(-4.90, -4.20)$ & $-4.50$ & \nodata \\
HD~1342 & 10.42 & 0.14 & 4.37 & 0.00 & 0.57 & $(-4.84, -4.14)$ & $-4.89$ & 0.042 \\
HD~103195 & \nodata & \nodata & 7.46 & 0.01 & 0.96 & $(-5.06, -4.48)$ & $-4.87$ & \nodata \\
HD~115405 & \nodata & \nodata & 6.33 & 0.01 & 0.85 & $(-4.80, -4.22)$ & $-3.82$ & \nodata \\
HD~131157 & 13.09 & 0.64 & 6.07 & 0.009 & 0.66 & $(-5.51, -4.81)$ & $-3.64$ & 0.124 \\
HD~205724 & \nodata & \nodata & 6.70 & 0.03 & 0.84 & $(-5.15, -4.57)$ & $-4.71$ & 0.042 \\
\enddata
\tablecomments{95\% confidence intervals for activity levels predicted from UV data. The sample consists of stars with $B-V$ and Ca~II measurements from \citet{sindex_duncan} that are not in the Hipparcos sample we used to calibrate the UV-activity relation. Where \citet{sindex_duncan} quoted multiple measurements, we selected one at random to ensure we did not suppress extreme or outlying measurements by averaging. Since the scatter of the measurements (column $\sigma_{\log{R^\prime_\textrm{HK}}}$) is small, the choice of epoch does not affect our results. Most predictions are within the 0.15-0.18~dex uncertainty of Equations~\ref{facfit} and \ref{nacfit}, but there are some exceptions, notably HD~131157. Aside from the two NGC~2632 members, none of these stars are discussed in the literature, so we have no information on whether they are interacting binaries, extreme metallicity stars, or otherwise unusual.}
\end{deluxetable}

\end{document}